# An overview of the use of alternative funding and contracting approaches relevant for agile software development: A systematic review of real-life experiences

Bertha Ngereja* and Magne Jørgensen

*bertha@simula.no; magnej@simula.no

Simula Metropolitan Centre for Digital Engineering (SimulaMet), Department of IT Management, Stensberggata 27, 0170 Oslo

## Abstract

Agile software development emphasizes flexibility and iterative processes, which may conflict with the more linear, rigid, and time-consuming traditional funding and contracting approaches. This review synthesizes real-life experiences of using alternative (non-traditional) contracting and funding approaches. The focus is on identifying approaches that align better with agile principles and understanding the motivations, benefits, and challenges these alternatives present. A systematic literature review was conducted in SCOPUS, Web of Science, and Google Scholar, where we identified 38 relevant peer-reviewed empirical studies from private and public sector contexts. Four alternative funding and four alternative contracting approaches were identified. Organizations were motivated to adopt these alternative approaches because traditional approaches often proved too rigid, conflicted with agile principles, hindered effective client-contractor collaboration, and limited profitability. The benefits of these alternatives included higher client satisfaction, reduced contractor risk, and more efficient resource utilization. Adopting alternative funding and contracting approaches may promote flexibility and efficiency in agile projects but also presents cultural and structural challenges, increases the risk of scope creep and analysis paralysis, and requires additional effort in terms of time and resources. The context of the organization matters highly in selecting a suitable approach, such as the organizational readiness in terms of its leaders, people, and systems. Thus, instead of wholly adopting alternative approaches and introducing changes abruptly, organizations may benefit from starting with hybrid approaches that balance flexibility and control and progressively transition to fully flexible approaches tailored to their needs

# 1. Introduction

The introduction of agile approaches has transformed the software development process (Medavarapu, 2023, Gupta and Gouttam, 2017). Integrating agile practices, such as flexibility in the early phases, increases both the effectiveness and efficiency of the front-end phase. In a B2B automotive industry where Scrum-Stage-Gate was adopted, a 30% increase in the speed of the development process was realized, suggesting that adopting agile practices in the early stages may increase agility, adaptability, and speed in the development phase (Cooper, 2016). Agile practices may provide higher benefits than traditional approaches by minimizing the time spent upfront in the planning phase (Moloto et al., 2021, Krehbiel and Miller, 2018, Conforto et al., 2016). Agile practices allow software projects to be planned with less detail, leaving room for flexibility to incorporate change in requirements, thus tend to display more success in terms of realizing benefits and controlling costs (Jorgensen, 2019).

Despite the increase in adopting agile approaches, tensions exist because organizations still need control and predictability (Lindskog, 2022). Partly because of such tensions, organizations choose to continue using traditional approaches for agile software development projects, which may have a negative effect on the outcome due to a lower ability to incorporate changes over time (Lappi and Aaltonen, 2017). Unlike working according to agile principles, traditional practices often focus on extensive upfront planning, requiring detailed documentation and specifications very early (Bustamante et al., 2016), resulting in long front-end phases.

To cope with continuous changes in technology and user needs, organizations in both the private and public sectors should aim for more adaptability and agility (Janssen and Van Der Voort, 2016). As documented by Jorgensen (2019), successful agile software development appears to be strongly connected with flexibility in deliveries and focus on learning through the delivery process rather than trying to fix deliveries and plans through a thorough front-end phase. Perhaps the most central characteristic of using agile methodology is the ability to adapt to changing needs, requirements, and benefits throughout the development process (Singh, 2021).

While agile practices are widely adopted during the development phase of software projects, they are often not well integrated into the front-end phase, such as securing project funding and determining contracting (procurement) approaches. These pre-development activities are still conducted using traditional methods, creating challenges and conflicts that can undermine agility during development. Kropp et al. (2020) found a positive relationship between the use of agile methods and higher team satisfaction and further suggested that employing traditional, plan-based approaches in agile software development may negatively affect project outcomes. This is supported by a recent study on agile SMEs, which found that agile practices are largely confined to the development team during the development phase, while supporting functions, such as budget management, continue to follow traditional methods, resulting in overall project delays (Henriquez et al., 2023).

This review explores studies providing practical experiences of alternative[1] (non-traditional) funding and contracting (procurement) approaches. The term "alternative" refers to practices that are different from traditional ones and offer flexible options that align with agile principles. In this paper, the term alternative funding refers to other existing options different from obtaining all the funding upfront, and alternative contracting refers to other types of contracts different from fixed-cost contracts.

The New South Wales (NSW) government in Australia has successfully applied an alternative funding approach through The Digital Restart Fund[2] and has already reported success stories[3]. To the best of our knowledge, the Digital Restart Fund, with its connected portfolio and management processes, contains the most extensive alternative funding approach for agile software development at the governmental level. Established in 2019 and still in its early stages, it illustrates the potential behind adopting alternative funding approaches. The fund aims to make digitalization projects' access to funds quicker (within 9 weeks) and more flexible (step-wise through seed or experimentation and scale funding). While not included in the reviewed papers, since the experience is not reported in any peer-reviewed research paper, the positive experiences support the importance of looking into the experiences of changing from traditional to more agile types of funding approaches.

In this review, we briefly describe the alternative approaches with empirical evidence, the reasons that motivate organizations to seek alternative approaches, and the real-life experiences (benefits and challenges) of organizations that have adopted these alternatives. Although both funding and contracting are financial mechanisms, they serve different purposes. Funding provides financial resources to achieve a goal. At the same time, contracting involves a client and service provider agreement for specific services in exchange for payment. This review examines alternative funding and contracting approaches separately because, despite being somewhat related, if an alternative funding approach is adopted, it may influence the choice of a contracting approach for a project. For example, if funding is provided step-wise based on value performance, then it would demand an alternative contracting approach, such as payment per sprint. However, there may also be cases where an alternative funding approach may be used with a traditional contracting approach, such as adopting a step-wise funding approach but agreeing on a fixed cost contract beforehand, allowing releasing of a fixed amount of funds at agreed milestones[4]. This review does not explore how alternative funding and contracting approaches complement one another; instead, it provides a comprehensive understanding of their characteristics, benefits, and challenges to equip practitioners with the insights needed for their effective adoption.

The rest of the article is organized as follows: Section 2 briefly presents related studies on alternative funding and contracting approaches and the research questions to be addressed. Section 3 presents the literature review methodology used. Section 4 presents the findings from the literature review. Section 5 discusses the findings and their limitations. Finally, section 6 concludes the review and presents future research opportunities.

[1] Our use of the term 'alternative' in this paper is loosely defined to include funding and contracting practices that align with agile principles, particularly in speeding up the front-end phase and providing greater flexibility during software development.
[2] https://www.digital.nsw.gov.au/funding/digital-restart-fund/about-the-fund
[3] https://www.digital.nsw.gov.au/article/digital-restart-fund-our-first-year?
[4] https://www.dau.edu/blogs/dod-releases-new-ot-guide

## 2. Related Studies

Funding authorities often prefer traditional funding and contracting approaches due to perceptions of better cost control (Lindskog, 2022). However, traditional approaches often require extensive upfront analysis, detailing of planned activities and bureaucratic processes (Bustamante et al., 2016). Traditional approaches are based on well-defined scope, cost, and schedule, which may to some extent conflict with the dynamic nature of agile projects, necessitating adaptations to accommodate their unique characteristics (Cao et al., 2013). Table 1 highlights key differences between traditional and agile software development, suggesting that there may also be differences in their need for funding and contracting approaches. Several agile researchers agree that it is time to challenge traditional approaches (Atkinson and Benefield, 2013, Lindsjørn and Moustafa, 2018).

*Table 1: Differences between traditional and agile development (Source: Dybå & Dingsøyr, 2008)*

| **Management approach** | **Traditional development** | **Agile development** |
|---|---|---|
| **Management style** | Command and control | Leadership and collaboration |
| **Knowledge management** | Explicit | Tacit |
| **Communication** | Formal | Informal |
| **Development model** | Life cycle | Evolutionary-delivery |
| **Desired organizational structure** | Mechanistic, aimed at large organization | Organic, aimed at small and medium organizations |
| **Quality control** | Heavy planning, strict control, heavy testing | Continuous control of requirements, design, and solutions |

The Stacey matrix (Stacey, 1996) and the CYNEFIN framework (Kurtz and Snowden, 2003) provide valuable insights into this issue. The matrix proposed by Stacey (1996), enables navigating the concept of complexity by categorizing projects into simple, complicated, complex and chaotic. The matrix shows that traditional contracts effectively address simpler or complicated contexts which have well-understood or predictable challenges (Atkinson and Benefield, 2013). However, agile software projects often fall into complex contexts characterized by continuous and unpredictable changes (Atkinson and Benefield, 2013). Thus, insisting on a traditional way of thinking for software projects is not optimal.

The CYNEFIN framework divides decision-making into five contexts: simple, complicated, complex, chaotic, and disordered (Kurtz and Snowden, 2003). Knowing which context applies enables choosing the most appropriate response to a specific problem (Turino and Santoso, 2020). Software development problems often fall in the 'complex' context characterized by unclear or evolving cause-and-effect relationships. In such contexts, solving problems typically requires experimentation and iterative approaches rather than extensive upfront planning. Snowden (2005) suggests that projects within complex contexts, such as software development, demand different strategies and decision-making processes compared to problems in simpler contexts.

Agile software development requires iterative and flexible practices rather than extensive upfront analysis and rigid planning. Alternative funding and contracting approaches for agile projects have thus emerged to address these challenges. Several authors highlight the need for modifying funding and contracting approaches to better accommodate agile principles. Huang et al. (2022) recommend adjusting traditional contracts by introducing flexibility, such as fixed-cost contracts with renegotiable scope. Hybrid approaches combining both traditional and agile elements have also been proposed, allowing organizations to exploit established methods while exploring new practices simultaneously (Cram and Marabelli, 2018, Vijayasarathy and Butler, 2015). On the contrary, approaches like Beyond Budgeting (Bogsnes, 2016), Incremental funding (Cao et al., 2013) and Payment per sprint (Zijdemans and Stettina, 2014), are not simply modifications on traditional approaches but have quite different characteristics.

Ocampo et al. (2021) argue against the use of traditional contracts to address the needs of agile software development. In their recent mapping study, they provide a comprehensive overview of agile contracting practices within the public sector. They found that the Time and Materials (T&M) contracting strategy was the most widely adopted, followed by Target Cost contracts. Other price-based contract types such as Two-phase contracts, Advantage contracts and Progressive contracts, were mentioned less frequently. Additionally, collaborative agile contracts were identified but rarely mentioned, appearing only once each as Incremental Delivery Contract, Target Term Contract, Contract Without Payment, and Fixed price Contract per Work Unit (Ocampo et al., 2021). Thus, suggesting either low adoption or limited recognition of these terms. de Brock et al. (2024) similarly identified Fixed-price and T&M contracts as prevalent approaches for software development. Furthermore, they highlight that in such cases, the Fixed price approach is not adopted as it but often adapted to fit agile project deliverables.

Book et al. (2012) further emphasizes that traditional approaches like Fixed price contracts, are unfair because it is impossible to set a fixed price for an agile project. They propose a contracting approach for agile projects that has a built-in risk limitation mechanism for the contractor, and a built-in cost limitation mechanism for the client.

Despite these developments, comprehensive research on alternative funding and contracting approaches for agile contexts is limited. Research has seen an increasing number of organizations transitioning to the use approaches (Lohan et al., 2013, Zijdemans and Stettina, 2014), but there remains a lack of consensus in literature regarding their effectiveness and practical adoption. As noted in the literature review by Jørgensen (2024), agile principles typically have yet to reach the front-end phase, where traditional approaches, including funding and contracting approaches, are used. Although some practical experiences with alternative (non-traditional) approaches exist, they are limited. These existing insights are fragmented and dispersed across various contexts and terminologies, creating challenges for practitioners aiming to understand the benefits and potential drawbacks.

To our knowledge, no prior systematic review synthesizes practical experiences of adopting alternative funding and contracting approaches for agile software development projects, as indicated by Jørgensen (2024). A review of practical experiences with alternative funding and contracting approaches may, amongst others, support agile practitioners in software development contexts in finding the '*sweet spot'* where they will be able to gain a proper balance between the level of control to exert, i.e., cost management, and the level of flexibility to meet customer satisfaction, i.e., benefit recovery for agile software development projects. This review aims to answer the following research questions:

*(1) What alternative funding and contracting approaches are relevant for agile software development projects?*

*(2) What are the motivations for adopting alternative funding and contracting approaches?*

*(3) What are the reported experiences of using such approaches, i.e. benefits and challenges?*

# 3. Method

A systematic literature review should follow an explicitly described and well-defined process to ensure reliability and reproducibility. We followed the systematic literature review guidelines by (Kitchenham, 2004). The process began with developing a review protocol which was adhered to throughout the review process.

## 3.1 Formulating the search string

Prior to formulating the search string, a preliminary literature search was conducted to identify various terminologies used to refer to alternative funding and contracting approaches. During the preliminary search, broad terms related to the research topic were explored, such as "contract* AND agile"; "fund* AND agile"; "contract* AND flexible"; fund* AND flexible"; "contract* strategy"; and "fund* strategy"; "open-ended contract" in Scopus, Web of Science and Google Scholar. The preliminary search narrowed the focus by identifying more specific keywords, forming the basis of the search string for the systematic literature review.

The specific keywords identified from the preliminary search adopted were then carried forward and used to formulate the search string. These keywords were *pay-as-you-go, target cost, times and materials, beyond budgeting, design to cost, incremental funding, tranche funding, seed funding, and experimentation*. Synonyms were identified and added for each of the keywords. Boolean operator OR and wildcards (* and " ") were used to make the searches more exact. To ensure the inclusion of relevant papers, we searched for these specific keywords in the title, abstract and keywords in Scopus and Web of Science. In Google Scholar, these specific keywords were searched in all parts of the article, including the title, abstract, keywords, and document text.

## 3.2 Databases and search engine

Multiple search tools, i.e., two databases and one search engine, were used to avoid overlooking relevant publications (Ewald et al., 2022). The databases used were Scopus and Web of Science, and the search engine used was Google Scholar. Scopus and Web of Science were chosen because they are very reputable academic databases. Despite the debate on whether Google Scholar is a sufficient search engine for conducting a systematic literature review, research supports its use as a complimentary tool in addition to other trusted research databases (Haddaway et al., 2015, Gehanno et al., 2013), such as Scopus and Web of Science. Some studies have found the coverage of Google Scholar much higher than initially thought, in fact Gehanno et al. (2013) found evidence in support of using Google Scholar alone for conducting systematic reviews. In addition, Google Scholar enables searches in the papers' body texts, not only in the titles and the abstracts, as is the case for the other library databases. Thus, Google Scholar was added to complement Scopus and Web of Science to ensure improved coverage of relevant publications. However, for effectiveness purposes, only the results from the first ten pages were included, which aligns with other reviews using Google Scholar as a complementary search engine (Johnstone et al., 2020, Moradi et al., 2021).

## 3.3 Inclusion and exclusion criteria

The inclusion and exclusion criteria applied are presented in Table 2 with their respective justifications.

*Table 2: Inclusion and exclusion criteria*

| Inclusion criteria | | |
|---|---|---|
| **Parameter** | **Inclusion criteria** | **Justification** |
| **Publication year** | No limit | To capture as many relevant papers as possible and to observe the trend of adopting alternative approaches over the years. |
| **Document Type** | Peer-reviewed journal and conference articles | To ensure that all articles included in the review passed through a rigorous peer-review process. Grey literature was excluded because a preliminary search revealed that it lacked well-documented experiences substantial enough to be included in the study. |
| **Subject area** | Business, management, and accounting, Engineering, telecommunications, and software engineering, Computer science, Information systems | To cover projects and other activities where agile methodology is adopted and to include more relevant contexts where alternative funding and contracting approaches could be used. |
| **Language** | English | The authors use English thus to facilitate a complete and easy understanding of the contents of the papers. |

| Exclusion criteria | | |
|---|---|---|
| **Parameter** | **Exclusion criteria** | **Justification** |
| **The study in the paper** | Papers that report on mathematical models, algorithms, and laboratory experiments | They would not provide any value in understanding alternative approaches on a real-life basis. |
| **Nature of evidence** | There is no empirical evidence presented | Purely theoretical studies were excluded because the review aims to synthesize real-life experiences where alternative approaches have been implemented. |
| **The focus of the paper** | There is no discussion of alternative funding and contracting approaches | This ensured that the review remains focused on alternative funding and contracting approaches. |
| **Potential usefulness to agile projects** | Papers whose experiences cannot be transferred to agile projects | As the review aims to gather experiences relevant to agile projects, those experiences that cannot be transferred to agile projects were excluded. |

## 3.4 Search Results

The database searches were conducted in February 2024. The keywords formed the basis of the search string, which led to many hits before applying inclusion criteria. The process was carried out by two reviewers: one conducted the initial screening, while the second performed a quality check to ensure accuracy and replicability. The search yielded 82,644 hits from WoS, 353,511 from Scopus, and 1,194,578 and from Google Scholar. In this first step, the high volume was anticipated as the keywords used are common and widely applied across various disciplines. Consequently, the initial results covered a broad range of subject areas, document types, and languages.

To narrow the scope to relevant literature, we applied predefined inclusion criteria (Table 2). Specifically, we included only peer-reviewed records published in English in journals or conference proceedings, within the fields of business, management, accounting, engineering, telecommunications, software engineering, computer science, and information systems. This filtering reduced the total number of records to 7,704 from WoS and 9,946 from Scopus. For Google Scholar, we limited inclusion to the first ten pages of search results for each sub-search within the main string, yielding 700 records. This filtration reduced the total number of records to 18,333.

The 18,333 records were then downloaded as RIS files containing information about the title, author, year of publication, journal or conference name, and abstract were downloaded and imported into the referencing software EndNote. 11 duplicate records were identified and excluded, which led to 18,322 records.

We then conducted title and abstract screening using our predefined exclusion criteria (Table 2) on the remaining records. This screening enabled us to identify the different contexts our keywords are used in. For example, the keyword *Pay-As-You-Go* appeared in 856 records, but majority of the records used the term in unrelated contexts, including Pay-As-You-Go billing for electricity and mobile subscriptions and Pay-As-You-Go pension. Also, the keyword *experimentation* led to a significant number of papers that referred to *laboratory experiments, mathematical models, and algorithms*. A similar screening process was applied to all keywords to ensure they were used in the intended context relevant to alternative funding and contracting approaches. This stage led to the exclusion of 17,872 records. A further 319 records were excluded upon closer screening of titles and abstracts for not presenting empirical findings, which was an essential requirement for inclusion. This left 120 papers for full-text review (3 from WoS, 69 from Scopus and 48 from Google Scholar).

The full texts of these 120 papers were assessed in detail. Of these, 97 did not address our research questions, particularly in terms of providing real-world experiences with adopting alternative contracting and funding approaches and were thus excluded. The remaining 23 papers were deemed sufficiently relevant and included in the final synthesis.

An additional 11 papers were further identified through forward and backward snowballing. To further enhance our review and minimize the risk of missing key contributions, we supplemented the database search with expert recommendations, which led to adding four papers. These papers were carefully assessed against our predefined inclusion criteria before being incorporated into the study. This additional step strengthened our review process by ensuring that valuable insights are not overlooked and reducing publication bias, as highlighted by Kitchenham (2004). In total, 38 papers were included in the final analysis to derive the findings presented in this review. The full search process is depicted in Figure 1.

| | |
|---|---|
| **Step 1: Preliminary search** | *Broad terms related to the research topic were explored, such as "contract* AND agile"; "fund* AND agile"; "contract* AND flexible"; fund* AND flexible"; "contract* strategy"; and "fund* strategy"; "open-ended contract" in Scopus, Web of Science and Google Scholar. The preliminary search narrowed the focus by identifying more specific keywords, forming the basis of the search string for the next step* |
| **Step 2: Search strings** | *("pay as you go" OR PAYG OR "pay-as-you-go") OR ("target cost" OR "target-cost" OR "target-cost*" OR "target-cost*") OR ("tranche OR "incremental fund" OR seed fund*) OR (time and material* OR T&M) OR "beyond budget*" OR ("design to cost" OR DTC) OR (experimentation OR "agile trial*" OR "agile-trial*")* |
| **Step 3: Databases used** | Web of Science<br>SCOPUS<br>Google Scholar |
| **Step 4: Inclusion criteria** | 1. Language: English<br>2. Document type: Journals and conferences<br>3. Subject area: Business, management, and accounting, Engineering, telecommunications, and software engineering, Computer science, Information systems<br>→ Are the criteria met??<br>Yes → Papers included in the next step:<br>Web of Science (n =7,704)<br>SCOPUS (n =9,946)<br>Google Scholar (n =700)<br>Total = **18,333 papers**<br>No → Exclude |
| **Step 5a: Screening of papers** | RIS files were downloaded and imported into a referencing software called EndNote<br>• Duplicate records removed (n =11)<br>• The titles and abstracts for the remaining publications were screened and excluded if:<br>- Discuss mathematical models, algorithms and laboratory experiments (n =17,872)<br>- No empirical evidence (n = 319)<br>Papers included in the next step of review (full paper reading) = **120 papers (WoS =3; SCOPUS = 69; Google Scholar = 48)** |
| **Step 5b: Full paper reading** | Papers excluded if they lacked real experiences of alternative funding approaches (not relevant to RQs) (n = 97)<br><br>Papers included from primary sources (n = 23)<br>Papers included from secondary sources:<br>• Snowballing (n = 11)<br>• Recommendation (n = 4)<br><br>Papers included in the review = **38 papers** |

*Figure 1: The search process*

## 3.5 Quality assessment

We performed a quality assessment on the credibility and practical value of the selected papers as suggested by Kitchenham (2004). We developed the following set of generic questions to assess the selected papers:

- Is there a clear statement of the aims of the study?
- Is there an adequate description of the context in which the research was carried out?
- Was the research method appropriate to address the aims of the research?
- Was the data collected in a way that addressed the research issue?
- Was the data analysis sufficiently rigorous?
- Is there a clear statement of findings?
- Is the study of value for research or practice?

For each assessment questions, the scoring was adopted from Kitchenham and Brereton (2013), that is YES (score=1), NO (score=0) and PARTIALLY (score=0.5). Considering this scoring type, the maximum score for each publication was 7. The quality score range (QSR) was categorized as LOW (QSR <3), AVERAGE (3≤QSR≤5) and HIGH (5≤QSR≤7).

To assess the trustworthiness and impact of the selected studies, we evaluated the publication venues of the included papers. Established metrics were used to indicate the quality and standing of the journals and conferences in which these studies appeared. These metrics were not used as an exclusion criteria, but rather to provide additional context regarding the trustworthiness and impact of the selected papers. Specifically, four journal ranking systems and one conference ranking system were applied, as outlined below:

- Journal Impact Factor: Reflects the average number of citations received by articles published in a journal over a specific period. Journals with IF < 1 are considered *emerging*, those between 1–3 *moderately regarded*, IF > 3 *well-established* and IF > 5 are considered *leading* within their fields.
- Harzing's Journal Quality (AJG 2024): Ranks journals on a five-point scale from 4* to 1 indicating *world leading, top, highly regarded, well regarded* and *recognized* respectively
- ABDC Journal Quality List (2022): Classifies journals in four categories from A*, A, B, and C, corresponding to *leading, highly regarded, well regarded,* and *recognized*.
- The Scimago Journal & Country Rank (SJR) score: Categorizes journals in quartiles (Q1 to Q4), where Q1 represents the top 25% journals in a subject area and Q4 represents the bottom 25%.
- The CORE conference: Ranks conferences in four tiers from A* to C indicating *top tier, high quality, good quality* and *moderate quality.*

## 3.6 Data extraction

The aim of the review was to identify the experiences of using alternative approaches in practice. In searching for these experiences in the identified papers, a data collection form (*Appendix ii*) was created to guide the data extraction process. Data extraction was done independently by both reviewers, after which the data was discussed to ensure consensus. The dataset extracted from the selected studies is available as an Excel file at Zenodo (DOI: 10.5281/zenodo.15689796). The dataset includes the following fields: Title, Authors, Source (Journal/Conference), Year of Publication, Study Location, Unit of Analysis (Project/Organization), Sector, Research and Data Collection Methods, Sample Size, Study Context, Comments on the Use of Traditional Approaches, Type of Alternative Approach Discussed, Motivation for Adopting Alternative Approaches, and Challenges and Benefits of Adoption.

The extracted data was then coded. Related codes were grouped, forming four thematic groups. Thematic group 1 covers the types of alternative approaches, i.e., whether the results relate to funding or contracting approaches. Thematic group 2 covers the motivations for adopting these approaches. Thematic group 3 covers the positive experiences of adopting alternative approaches, while thematic group 4 covers the negative experiences (challenges). The thematic groups and the codes used for the topics within each thematic group are presented in Figure 2. The findings for each thematic group were synthesized.

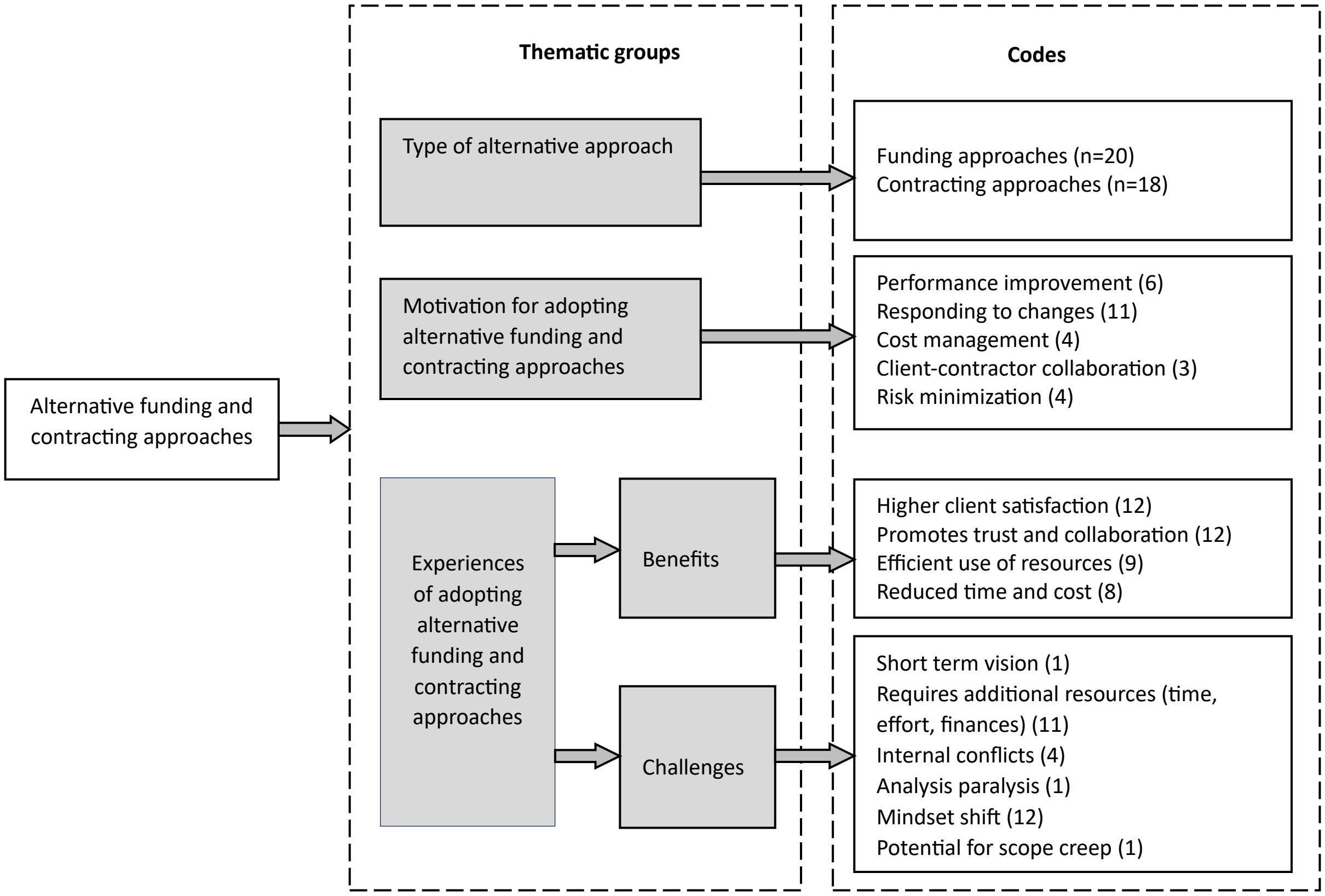

*Figure 2: Themes and codes from the selected papers*

# 4. FINDINGS

The review identified 38 studies, which included twenty-nine (76%) journals and nine (24%) conference papers. All selected papers scored high in the quality assessment (see *Appendix i*), providing confidence in the credibility and relevance of their findings. Furthermore, all selected papers were published in reputable, peer-reviewed journals or conference proceedings, which are ranked in one or more recognized academic rankings systems. For IF, most papers were published in journals with IF> 3, indicating that they are well-established and influential journals; for AJG, majority of papers are published in journals ranked 3 or above; for ABDC, majority are published in journals ranked A or A*; for SJR, majority of papers are published in the top 25% (Q1) journals; and the conference papers are from conferences ranking A or B, as shown in Figure 3. Being published in good to top tier quality conferences and highly regarded to world leading journals indicates that the selected papers have undergone a rigorous peer-review process. This enhances the trustworthiness and robustness of selected studies, thereby strengthening the reliability of the insights synthesized in this review.

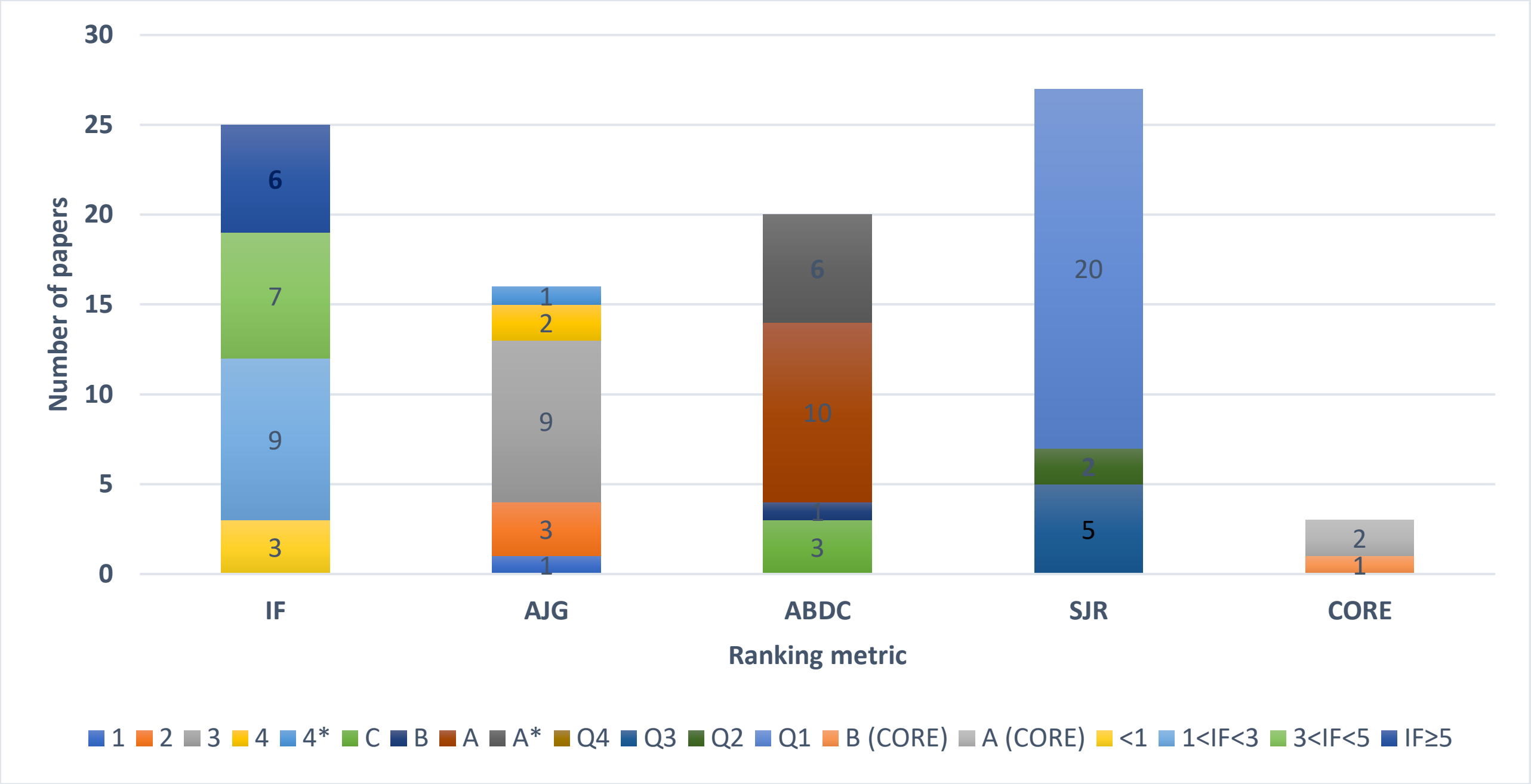

*Figure 3: Journal and conference rankings of selected papers*

No restrictions were placed on the publication year to capture as many relevant discussions on alternative funding and contracting approaches as possible. 3% of the papers are from the year 2000, while the rest (97%) span from early 2000 to 2022, as shown on Figure 4. This trend aligns with the rise of the Agile Manifesto in the early 2000s (Fowler and Highsmith, 2001), which highlighted the mismatch between agility and rigid traditional practices. The trend shows a slow, but steady increase in interest on this topic over time.

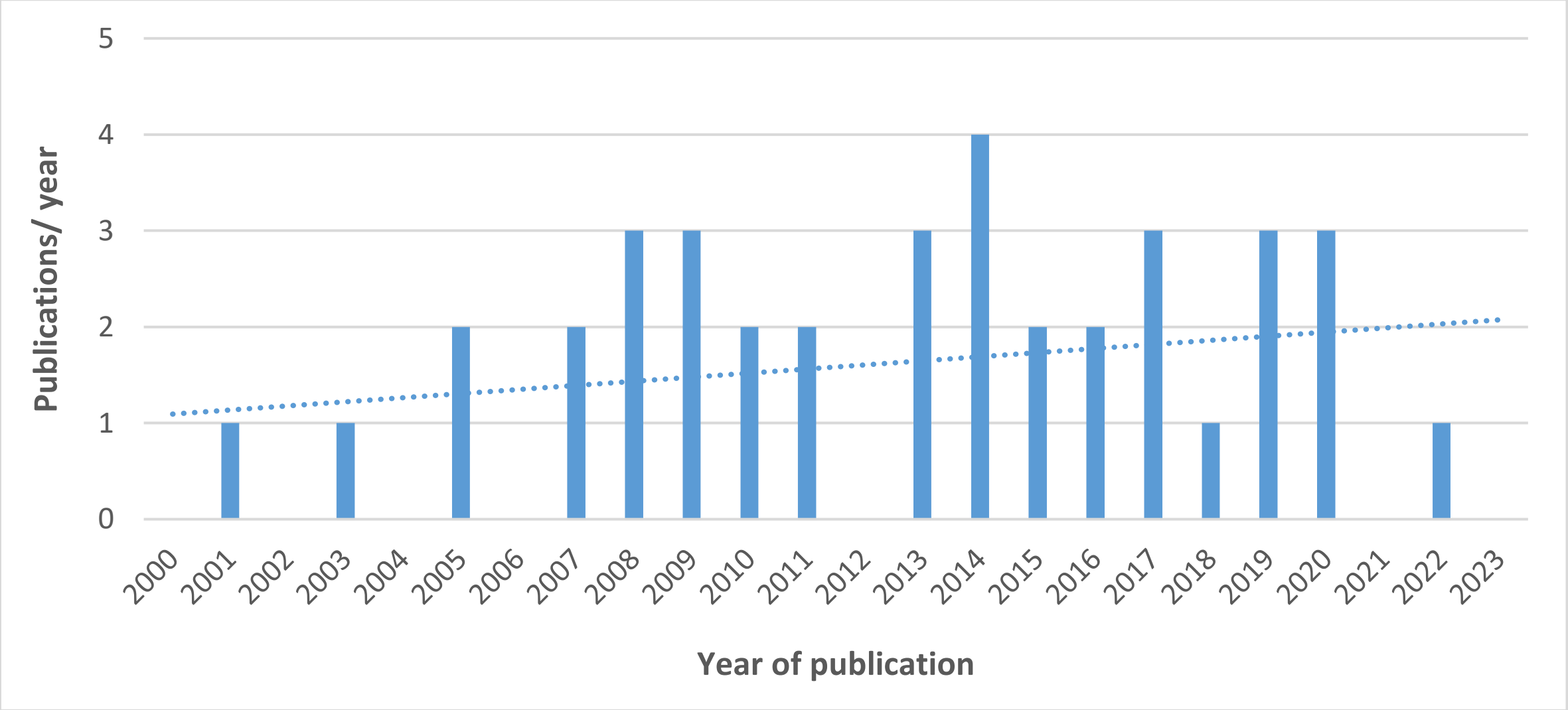

*Figure 4: Trend of publications per year*

The review found that most papers (76%) used interviews, observations, and/or archival documents for data collection. Furthermore, the identified papers covered diverse projects and organizations across various industries, most (45%) focused on agile software development. Other industries covered include construction (8%), manufacturing, specifically new product development (18%), hospitality industry (3%), oil and gas (3%), agriculture (3%), and housing (3%). 18% of the papers did not specify a context. Danoesastro et al. (2021) highlighted success stories of abandoning traditional budgeting in a Swedish bank and a Norwegian oil and gas company, demonstrating the applicability and benefits of alternative funding and contracting approaches across different sectors. Research also suggests that experiences of using alternative approaches, such as contracting, can be transferred from the software industry to other businesses, as highlighted in (Nuottila et al., 2015). We believe this can work both ways and that learning from different sectors beyond software can help deepen our understanding of how these approaches can benefit software development. For this reason, we include experiences from other industries where alternative approaches relevant to agile software development have been adopted. For a complete overview of these findings, see *Appendix iii*.

## 4.1 Descriptions of alternative funding and contracting approaches

We identified four funding and four contracting approaches that we considered alternatives to the traditional approaches and better aligned with agile principles, see Table 3.

### 4.1.1 Alternative funding approaches

Although budgeting is an important tool for management control, the research considers traditional funding to be outdated, especially in the context of how a project budget is established (Cardoş, 2014, Libby and Lindsay, 2010). This has led to much criticism and the development of new budgeting and funding concepts. Our review identified the actual use[5] of the following four alternative funding approaches: *Beyond budgeting, Step-wise funding, Pay-As-You-Go*, and *Funding for experimentation*. The categories are described below.

#### *a) Beyond Budgeting (BB)*

Beyond budgeting is an alternative funding approach that is not merely about getting rid of budgets, but advocates for forming more agile and human-centered organizations (Bogsnes, 2016). Beyond Budgeting was developed in the late 1990s as an alternative management model based on practices in a few EU and US organizations (Balaban and Đurašković, 2021). By the early 2000s, it gained attention alongside the rise of agile software development, with which it shares some similar principles (Lohan, 2013). The distinguishing feature of this approach is that it enables continuous updating of financial plans with a strong emphasis on long-term value creation rather than short-term financial targets. This mindset makes it possible for organizations to continuously update financial plans based on actual performance and changing business conditions.

Beyond Budgeting is the most frequent (24%) discussed alternative funding approach in our review, discussed in nine (24%) of reviewed papers (Alsharari, 2020, Broccardo et al., 2016, Aksom, 2019, Becker,

[5] As described earlier, we only include funding and contracting approaches where actual use has been reported in the research literature. We use the funding and contracting terms used in the papers in our discussion of the approaches.

2014, Bourmistrov and Kaarboe, 2013, Ostergren and Stensaker, 2011, Lohan et al., 2013, Nikolaj Bukh and Sandalgaard, 2014, Becker et al., 2020). Its popularity in the 2000s may be linked to its shared principles with agile approaches, which were also gaining traction then.

From the review, organizations adopted Beyond Budgeting by replacing traditional annual fixed budgeting with rolling forecasts (Alsharari, 2020, Broccardo et al., 2016, Aksom, 2019, Becker, 2014, Bourmistrov and Kaarboe, 2013, Ostergren and Stensaker, 2011, Nikolaj Bukh and Sandalgaard, 2014, Lohan et al., 2013, Becker et al., 2020). Rolling forecasts are a *"just-in-time process that focuses on strategy, threats, and opportunities and allows firms to allocate or withhold resources quickly and efficiently"* (Lorain, 2010).

An organization that adopted Beyond Budgeting, discussed in Nikolaj Bukh and Sandalgaard (2014) shifted from traditional bottom-up budgeting to top-down targets for business units. These targets were fixed for the year and focused on achieving financial goals like profit margins. The budget structure was simplified by removing many line items to a less detailed one to enable them to reach high-level financial goals rather than micromanaging every detail. Although the targets were fixed, managers had autonomy in achieving the goals, such as finding innovative ways to meet the targets despite any changes in external conditions.

Another organization discussed in Ostergren and Stensaker (2011) that initially had a traditional budgeting process that combined target setting, planning, and resource allocation adopted Beyond Budgeting by eliminating budgets and breaking down their combined processes into individual processes. The breakdown enabled shifting from focusing on detailed and fixed financial targets to more strategic and high-level goals, giving managers more room to adapt to changing conditions.

#### *b) Step-wise funding*

Step-wise funding involves making funds available step-by-step as the project progresses. What sets this approach apart from other approaches that make funds available in steps is that the decision to provide the additional funds is made depending on project evaluations conducted within a specific agreed time period, for example, quarterly or monthly, or at any time period as needed. This approach ensures that customer-valued features are delivered in sequenced chunks to optimize the project's net present value (Denne and Cleland-Huang, 2004).

The review found several variations of Step-wise funding, discussed in seven papers (18%) (Cao et al., 2013, Thomas and Baker, 2008, Brander Brown and Atkinson, 2001, Frow et al., 2010, Ekholm and Wallin, 2011, Berland et al., 2018, Sirkiä and Laanti, 2015). The reviewed papers use different terms to denote a funding approach that allows for frequent revisions and adjustments, i.e., Step-wise funding, instead of waiting for a predetermined budget revision date as in traditional budgeting. In the review, one paper used the term *Incremental funding* (Cao et al., 2013); one used *Flexible funding* (Thomas and Baker, 2008); one used *Continuous budgeting* (Frow et al., 2010); one used *Non-traditional budgeting* (Berland et al., 2018), and two used *Flexible budgeting* (Brander Brown and Atkinson, 2001, Ekholm and Wallin, 2011). Although the terms differ, they all have characteristics of a Step-wise funding approach.

The Step-wise approach discussed by Cao et al. (2013) started with a 3-month startup period to organize the development teams and the governance structure. After the startup period, funding decisions were made and reviewed every quarter. Throughout the development duration, the client and contractor worked closely to enable the client to specify the immediate scope of the project over the subsequent five or six development cycles, and funding was made available only for those activities. The step-wise approach discussed in Thomas and Baker (2008) involved funding projects in a portfolio where instead of planning projects annually, it was done in shorter cycles of three months where the scope, budget and priorities could be adjusted. The projects were funded based on their value (criticality) to the business, and resource availability was decided as required.

The Step-wise funding approach discussed in Brander Brown and Atkinson (2001) and (Ekholm and Wallin, 2011) involved supplementing the annual budget plans with frequently revised forecasting and resource allocation updates. However, both papers do not specify the exact duration between these revisions. The step-wise approach discussed in Berland et al. (2018) replaced detailed budgets with partial budgets, which were updated quarterly based on performance. Similarly, in Frow et al. (2010)Step-wise funding involved creating budgets or, as the organization called it, "control plans," which were reviewed monthly. The organization also had an annual outlook (forecast), which was revised every quarter, and necessary adjustments were made. They also had monthly budget reviews which informed the quarterly outlook review, which then informed the annual outlook with the overview for full-year achievement.

*Adaptive finance and control* is a term used in one paper (Sirkiä and Laanti, 2015) to denote a variant of Step-wise funding. Before introducing this approach, the organization required detailed cost estimates for up to 18 months, and the project budgets were fixed and required pre-approval for changes. This alternative approach involved getting rid of the long-term cost estimates and, instead, short-term cost estimates were used, making the estimates more accurate and adaptable. Projects were planned in smaller batches (up to 3 epics), and each project was prioritized based on whether it progressed well. In addition, this approach abandoned the project-level budgets that needed continuous updating and re-approvals, which meant there were no project overruns and no variance analysis against original plans. Business plans were created based on minimum requirements, and nice-to-have were considered as extra. In addition, this approach had a specific lump sum to support innovation and quality. Furthermore, this alternative approach was created by combining principles of agile, lean and beyond budgeting. The practice of integrating Lean and Agile in software development is not new. Existing research shows that Lean can improve Agile processes and vice versa (Wang, 2011). Similarly, Suomalainen et al. (2015) reported that more organizations are combining agile and lean practices to leverage the flexibility of agile in dynamic and turbulent environments.

### *c) Pay-As-You-Go based funding*

The Pay-As-You-Go (PAYG) approach is discussed in one paper (3%) (Cao et al., 2013). This approach involves breaking down the project into several mini-projects instead of treating the project as one big task. Each mini-project then focuses on developing a few specific features and lasts four to six two-week development cycles. The costs and schedules are agreed upon in advance for each mini-project, and the client funds each mini-project one phase at a time instead of paying for the entire project upfront. This approach shares characteristics with Step-wise funding but differs in, amongst others, that the project is considered as a set of mini-projects rather than a step-wise execution of one project. This may lead to even higher flexibility in the project execution than for the Step-wise approach.

### *d) Funding of experimentation*

Funding of experimentation involves providing funds to evaluate ideas, hypotheses or concepts at a smaller scale, for example, developing prototypes of minimal viable products (MVPs) before full-scale implementation. Instead of going directly into full-scale procurement without knowing whether it would work, experimentation enables testing new software solutions in real-life conditions before implementation. Alternatives of funding of experimentation are discussed in three (8%) of the reviewed papers (Zorzetti et al., 2022, Soe and Drechsler, 2018, Ramakrishnan, 2009). This funding approach may be considered a type of Step-wise funding if it is done in a “pipeline” such that the experimentation leads to a scaled-up project if successful and if the funding for scaling up originates from the same source as the funding for experimentation.

In Zorzetti et al. (2022), funding was provided for a dedicated experimental research lab. It took three months to set up the lab configuration/room renovation and nine months to ensure it worked as needed. In this case, experimentation was to be integrated as part of the process to ensure that the products that will be produced match customer satisfaction before they reach the market. In Soe and Drechsler (2018), the funding for experimentation was set at Euros 67,000 and was opened to the public to submit proposals for conducting the trials (experimentation). In the process, two innovation labs won and conducted the agile trials. Another funding amounting to Euros 1,000,000 was set aside for the subsequent phases (real pilots) and solutions, and the plan was to procure the services separately. In Ramakrishnan (2009), funding was provided for an experimental project to introduce structured software engineering processes to student teams. After the experimental project proved to have a positive impact, the university secured more funding to open a dedicated lab to scale it up.

### 4.1.2 Alternative Contracting Approaches

For this review, we assumed all contracting approaches that are not fixed-price, fixed-scope belong to the category of alternative contracting approaches. With this assumption, we identified four alternative contracting approaches: *Target Costing, Payment per sprint/ milestone, Two-phase contract, and Times and materials (T&M)*. The relevant papers and the approaches are briefly described below.

#### *a) Target cost*

Target costing is an approach that focuses on setting the quality, price, reliability, delivery terms, and targets during product planning and development to meet customer’s perceived needs and interest (Melo et al., 2014). What sets Target costing apart from other approaches is its emphasis on shared responsibility and risk sharing between the client and contractor. In our review, Target cost contract is discussed in eleven (29%) of the reviewed papers (Afonso et al., 2008, Badenfelt, 2007, Hattami et al., 2020, Ballard and Rybkowski, 2009, Baharudin and Jusoh, 2015, Ax et al., 2008, Dekker and Smidt, 2003, Jacomit and Granja, 2011, Jørgensen, 2017, Eckfeldt et al., 2005, Cooper and Chew, 1996).

Target cost contracts have been around since the 1930s (Feil et al., 2004). Japanese companies adopted it in the 1960s (Sharaf-Addin et al., 2014), and since then, it has gained relevance across various industries, including high-tech consumer goods and project-based sectors like manufacturing and construction (Zimina et al., 2012). The long history and widespread use of Target costing across various sectors may explain why it is a frequently reported alternative contracting approach in our review.

There may be differences in how Target cost contracts are implemented to suit specific project characteristics. In the Target cost contract discussed in Badenfelt (2007), the agreement was that the contractor would share 30% of any savings below the target cost but also bear 30% of the costs if the price exceeded the target. There was a ceiling beyond which the contractor would not be compensated for additional overruns. A manufacturing organization discussed in Afonso et al. (2008) adopted Target Costing in iterations by adjusting product design, components, and manufacturing processes to ensure that the final product could meet the target cost. Instead of controlling costs during production, the company shifted its focus to managing costs early in the design stage. This proactive approach allowed them to design cost-competitive products before manufacturing even began.

In a construction project discussed in Ballard and Rybkowski (2009), the target cost was set early after the client and key team members had validated the business plan. Instead of designing the building and pricing it later, as done in the traditional approach, the project team started with a fixed target cost and designed the building within that cost. Multiple design alternatives were explored and tested to find the most cost-effective solutions. In Baharudin and Jusoh (2015), an automotive organization adjusted its designs and processes based on the target cost. They shifted the cost control focus to the early stages of product development, helping the organization avoid expensive changes later in production.

#### *b) Payment per sprint/milestone*

Agile projects often use sprints to deliver functionality in an iterative manner and within a specific period (Sharma and Kumar, 2019). The review found that different terminologies may be used to denote a payment per sprint/milestone contracting strategy. Payment per sprint/milestone involves an agreement between a contractor and a client where the contractor is paid after an agreed sprint or milestone is completed. In such contracts, the client pays after ensuring that the contractor has met the specifications. This contract also facilitates ongoing feedback through sprint reviews to ensure the project stays on track. Payment per sprint/milestone is discussed in two (5%) papers (Thorup and Jensen, 2009, Zijdemans and Stettina, 2014).

While this method may appear similar to Step-wise funding, it is different. The key difference is that Payment per sprint/milestone is based on the completion of a specific sprint, whereas Step-wise funding involves evaluations conducted at predefined intervals (e.g., weekly, monthly, or quarterly), where an assessment of the overall progress is made rather than just the outcome of a single sprint. Additionally, in Step-wise approach, funding is typically applied at the project level, whereas Payment per sprint/milestone funding is applied at the sprint level within the project.

Our review found that the term Collaborative Agile Contract is used to denote Payment per sprint contracts (Thorup and Jensen, 2009, Zijdemans and Stettina, 2014). In both papers, the term collaborative agile contract describes a joint understanding between the client and supplier where both parties agree on shared responsibilities and mutual expectations in the early stages. Instead of traditional one-time payments, the collaborative agile contract in Zijdemans and Stettina (2014) introduced payment per sprint, where the client would pay for each completed phase or increment of the project based on deliverables rather than a fixed scope or budget. Both parties could renegotiate priorities and add or remove features during each sprint. The contract emphasized trust between both parties, allowing the supplier more freedom in determining how the work was done.

In Thorup and Jensen (2009), the collaborative agile contract focused on outlining only a broad vision of the project scope and allowed details to be added incrementally. The payment was linked to the completion of specific milestones. Each milestone was tied to the client's software deployment, ensuring payment was made for actual, usable outcomes rather than pre-set deliverables. The contract incentivized the supplier and customer to finish the project quickly and with the right amount of functionality. The contract featured an hourly rate for ongoing work, which was lower than standard rates. In addition, there was a completion bonus upon reaching the agreed-upon milestones.

In Zijdemans and Stettina (2014), a fixed amount per sprint was established instead of setting one for the entire project. Each sprint lasted between two to four weeks. At the end of each sprint, the client would pay based on the agreed price per sprint, regardless of the exact hours spent. The organization billed the client immediately after each sprint was completed and accepted. The payments followed the agile delivery cycles, meaning the client paid as they received value from each sprint's work. A sprint review was conducted at the end of each sprint to demonstrate the work completed. The client had the opportunity to provide feedback and request adjustments for the following sprint, further ensuring alignment with expectations. The client could continue purchasing more sprints based on their satisfaction with the work. This created an ongoing cycle of feedback, deliverables, and payment, making it easier for the client to manage the evolving scope and budget.

#### *c) Two-Phase Contract*

The experience of adopting a Two-Phase Contract is discussed in one paper (3%) Zijdemans and Stettina (2014). This approach is different from other identified approaches in that instead of having all activities in a single contract as done in fixed-cost approaches, it has separate agreements for the planning and development phases. In a software project discussed in Zijdemans and Stettina (2014) where a Two-phase contract was adopted, Phase 1 focused on requirements gathering, discovery, and initial planning. During this phase, the project team worked closely with the client to define high-level requirements, objectives, and potential risks. This phase was time-boxed and billed based on effort (e.g., hourly or daily rates). The goal was to outline the project scope, define the backlog of features, and provide a more accurate estimation of the overall cost and timeline for the development phase.

Phase 2 operated under a flexible payment model, typically with payment per sprint or feature delivery, depending on the progress. The development phase allowed for adaptive changes and prioritized features based on ongoing feedback from the client. Deliverables were reviewed at the end of each sprint or iteration, and the client could decide to continue or adjust the project based on the evolving needs. This allowed the client and supplier to refine the project goals and estimates after understanding the requirements in Phase 1. The second phase allowed ongoing adjustments to the project scope and priorities, accommodating the evolving nature of agile projects. The client had more control over the budget and could reprioritize features mid-project.

#### *d) Time and Materials (T&M)*

Materials (T&M) is a contract where the client pays the contractor based on the actual time and materials spent on the project. While it may be argued that this is a traditional type of contract, in the sense that it is already in widespread use, we include it here since it may be implemented to fit agile principles well. In many ways, it may be viewed as the most flexible of all contract types. It does, for example, not require a complete up-front requirement specification or a pre-determined budget. What makes this approach different from other approaches is that payment is made based on actual effort in terms of time spent and materials used. Time and Materials (T&M) contracts are discussed in five (11%) papers (Fink and Lichtenstein, 2014, Gopal and Koka, 2010, Jørgensen et al., 2017, Persson et al., 2022, Jørgensen and Grov, 2021). Time and

In the T&M contract discussed in Gopal and Koka (2010), the payments were based on the actual time spent and materials used by the contractor in developing the software, including the hourly rates for the development team and reimbursement for any materials or resources used. The T&M contract allowed for scope changes and project adjustments without the need for renegotiation or fixed deadlines. In the T&M contract discussed in Fink and Lichtenstein (2014), the client paid for the actual time and effort the development team spent rather than a predefined cost. This allowed the development team to be paid hourly or for the total hours worked, with materials used also reimbursed separately.

In the T&M contract discussed in Jørgensen et al. (2017), the client agreed to pay for the actual hours worked by the development team and other incurred costs instead of having a pre-defined scope of work. The payment rates were set for different roles or skills in the development team. The contract allowed for flexible scope changes throughout the project. Changes could be made to requirements without the need for costly contract renegotiations. The client could provide feedback on each increment or sprint, adjust priorities, and monitor the progress, thus keeping the project aligned with their evolving needs. The team and client could adjust the workload and resources based on project needs, allowing for ongoing reallocation of resources to match the highest-priority tasks. In addition to the basic features of a T&M contract, the client can add some additional features to ensure that the contractor is held accountable. The T&M contract discussed in Persson et al. (2022), had penalties such as a fee if a team member was substituted, a week of mandatory team training, and a four-month trial period during which the client could cancel the contract if they were unsatisfied with how the relationship developed.

*Table 3: Alternative funding and contracting approaches*

| | Alternative approaches | Key characteristics | Papers |
|---|---|---|---|
| **Funding approaches** | **Beyond budgeting (BB)** | • Continuous update of financial plans<br>• Focus on long-term value creation instead of short-term financial targets | (Alsharari, 2020, Broccardo et al., 2016, Aksom, 2019, Becker, 2014, Bourmistrov and Kaarboe, 2013, Ostergren and Stensaker, 2011, Lohan et al., 2013, Nikolaj Bukh and Sandalgaard, 2014, Becker et al., 2020) |
| | **Step-wise funding** | • Funding decisions are made frequently (i.e., quarterly, monthly etc.,)<br>• The next funding in the pipeline depends on the previous performance | (Brander Brown and Atkinson, 2001, Frow et al., 2010, Ekholm and Wallin, 2011, Berland et al., 2018, Sirkiä and Laanti, 2015, Cao et al., 2013, Thomas and Baker, 2008) |
| | **Pay-as-you-go based funding (PAYG)** | • Fund for each mini-project is agreed upfront<br>• Funds made available at the start of each mini-project | (Cao et al., 2013) |
| | **Funding of experimentation (FoE)** | • Funding provided for agile trials at a small scale before scaling up | (Zorzetti et al., 2022, Ramakrishnan, 2009, Soe and Drechsler, 2018) |
| **Contracting approaches** | **Target cost (TC)** | • Risk is shared between client and contractor<br>• Establish the target "allowable" cost early and try to align actual costs to it | (Afonso et al., 2008, Badenfelt, 2007, Hattami et al., 2020, Ballard and Rybkowski, 2009, Baharudin and Jusoh, 2015, Ax et al., 2008), (Dekker and Smidt, 2003, Jacomit and Granja, 2011, Jørgensen, 2017, Eckfeldt et al., 2005, Cooper and Chew, 1996) |
| | **Payment per sprint/milestone (PPS)** | • Payment per successful sprint or increment<br>• Risk is shared between client and contractor | (Thorup and Jensen, 2009, Zijdemans and Stettina, 2014) |
| | **Two-phase contract** | • Different contracts for planning/ design and development phases | (Zijdemans and Stettina, 2014) |
| | **Time and Materials (T&M)** | • Payment based on actual effort, i.e., time and materials used | (Fink and Lichtenstein, 2014, Gopal and Koka, 2010, Jørgensen et al., 2017, Persson et al., 2022, Jørgensen and Grov, 2021) |

The review found that although the experiences with adoption were globally dispersed, Europe accounted for the most diverse and extensive adoption, representing 59% of reviewed papers. We noted that 61% of the studies from Europe originated from the Nordic countries, specifically Norway, Sweden, Finland and Denmark. As illustrated in Figure 5, Europe leads in the adoption of six out of eight identified alternative approaches. The only approach not represented in Europe is Pay-As-You-Go, but the remaining seven all have experiences from the region. Some approaches have exclusive experiences in Europe and were not found in any other locations. Following Europe, Asia is represented through three alternative approaches: Beyond budgeting, Target cost and Time and Materials. North America is represented in two alternative approaches: Step-wise funding and Target cost. This distribution highlights the leading role of Europe, especially Nordic countries in experimenting with and researching a wider range of alternative funding and contracting practices.

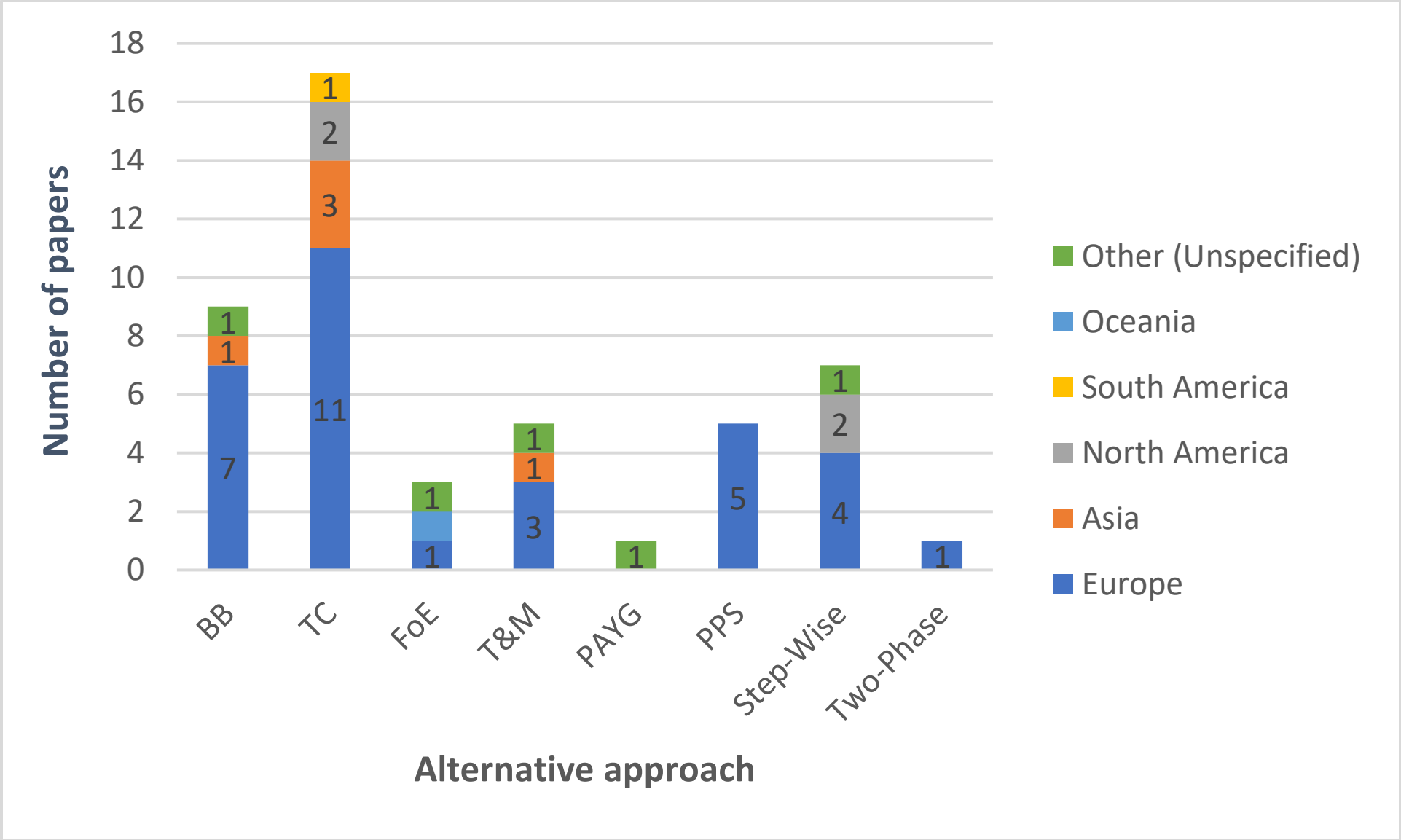

*Figure 5: Adoption of alternative approaches by location*

The review also found that thirty-two (84%) focused on the private sector, four (11%) on the public sector, and two (5%) on both sectors. This shows that most approaches are predominantly adopted in the private sector, while the public sector shows a limited presence. In the public sector, the few experiences identified were primarily associated with Time and materials and Funding of experimentation approaches. Some approaches, such as Step-wise funding, did not have any experiences from public sector, while others including Payment per sprint, Pay-As-You-Go and Two -phase contract appeared rarely and only in private sector. These findings are depicted in Figure 6. Our finding are consistent with existing research suggesting a slow adoption of alternative funding and contracting approaches in the rigid structure of the public sector (Nuottila et al., 2016, Baxter et al., 2023, Bogdanova et al., 2020). McCarthy and Lane (2009) noted that although the public sector had begun moving toward decentralizing decision-making, most critical financial and management decisions remained hierarchical and followed traditional top-down strategies.

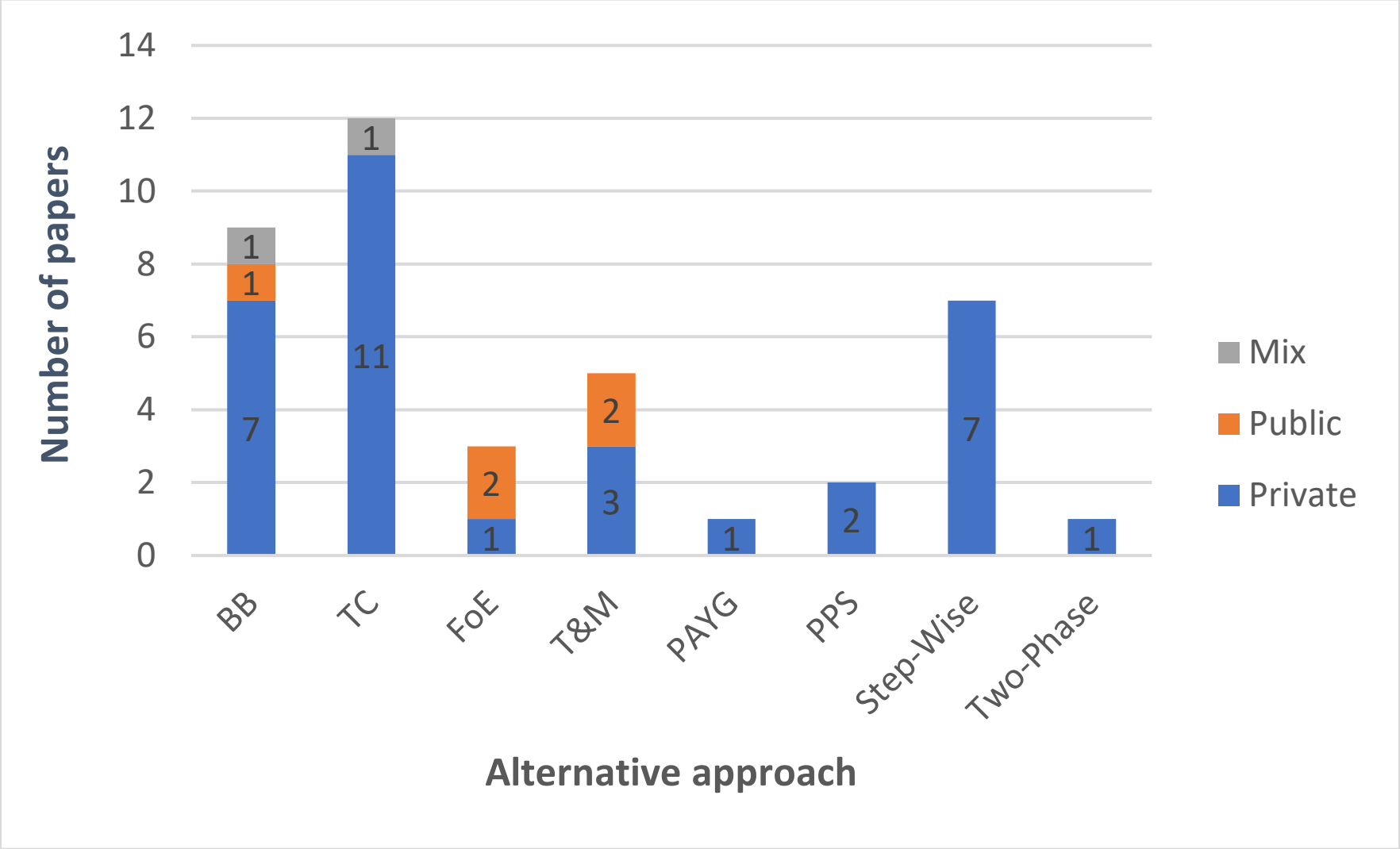

*Figure 6: Adoption of alternative approaches by sector*

## 4.2 Motivation for adopting alternative funding and contracting approaches

Traditional funding and contracting approaches may offer limited flexibility, which conflicts with agile principles. Making early commitments on scope and cost may be counterproductive for agile software development projects (Cooper, 2016). The traditional budgeting approach has been criticized since the 1960s, and although numerous attempts to replace it have been unsuccessful, it remains widely used despite its evident shortcomings (Lohan, 2013). Regardless of whether the approach is for funding or contracting software projects, our findings support existing studies that argue against the continued use of traditional approaches in dynamically changing environments. One reviewed paper by Sirkiä and Laanti (2015) suggest discontinuing the use of funding approaches designed for the industrial age, as they conflict with lean and agile principles.

Our findings suggest that the primary motivation for organizations in various contexts to seek alternative funding and contracting approaches is to overcome the limitations of traditional approaches, which are identified below;

- Five (13%) of reviewed papers suggest their incompatibility with agile principles (Alsharari, 2020, Aksom, 2019, Becker, 2014, Thorup and Jensen, 2009, Zijdemans and Stettina, 2014);
- Eight (21%) of reviewed papers suggest that they are too rigid and time-consuming in nature (Bourmistrov and Kaarboe, 2013, Ostergren and Stensaker, 2011, Becker et al., 2020, Lohan et al., 2013, Brander Brown and Atkinson, 2001, Cao et al., 2013, Frow et al., 2010, Ekholm and Wallin, 2011);
- Five (13%) of reviewed papers suggest they lack transparency, which hinders client-contractor collaboration (Badenfelt, 2007, Ballard and Rybkowski, 2009, Badenfelt, 2008, Eckfeldt et al., 2005, Jørgensen et al., 2017) and
- Three (5%) of reviewed papers suggest that they hinder creativity and innovation, which may lead to reduced profit margins (Hattami et al., 2020, Cooper and Chew, 1996, Zorzetti et al., 2022).

These findings align with existing studies, such as the systematic mapping study by Ocampo et al. (2021), which showed that organizations and software projects are transitioning from traditional contracting approaches to more flexible ones. Book et al. (2012) also oppose traditional contracting approaches due to their unfair risk distribution between involved parties. However, it is important to highlight that our findings are synthesized from experiences both in software development projects and other contexts, demonstrating that traditional approaches are unsuitable for agile software development projects and other projects and organizations operating in dynamic environments.

This review found the need for flexibility is the primary motivator for shifting from traditional to alternative funding and contracting approaches. Other motivators include the need to build collaboration and trust between clients and contractors, the opportunity to defer decision-making, which speeds up the planning phase, and the need to optimize the allocation and usage of resources. The specific motivations for adopting each approach, except for Pay-As-You-Go-based funding, which was not explicitly described, are presented below.

### 4.2.1 Motivations for adopting alternative funding approaches

The reviewed papers showed that Beyond Budgeting has been adopted across small and large organizations and sectors such as oil and gas, hospitality, agriculture, software development, and construction. In five (13%) of the papers, it was found that the primary motivation for adopting Beyond Budgeting was to improve performance and achieve organizational goals (Bourmistrov and Kaarboe, 2013, Broccardo et al., 2016, Nikolaj Bukh and Sandalgaard, 2014, Ostergren and Stensaker, 2011, Becker et al., 2020). An oil and gas company discussed in Bourmistrov and Kaarboe (2013) adopted Beyond Budgeting because it needed to increase its productivity, and it found that having annual (traditional) budgets was an excuse not to start more projects. As found in Broccardo et al. (2016), traditional budgets can lead to gaming the system, such as spending more money at the end of the budget period, estimating more costs, or even negotiating easier targets, which may have a negative impact on performance. Other existing studies like that of Alhalawi and Dammak (2024) and Player (2003) support this finding by emphasizing performance improvement as a key driver for adopting Beyond Budgeting. Our review also found that organizations are driven to adopt Beyond Budgeting aiming to manage costs in software development effectively (Lohan et al., 2013), and responding quickly to changes to boost productivity and reduce risks (Broccardo et al., 2016, Becker, 2014, Bourmistrov and Kaarboe, 2013).

Step-wise funding approaches have been applied in diverse contexts. In five (13%) of the papers, the adoption of Step-wise funding was driven by the need for flexibility to quickly respond to changes to support the dynamic and iterative nature of agile methodology and to manage ongoing uncertainties in changing environments (Cao et al., 2013, Frow et al., 2010, Ekholm and Wallin, 2011, Brander Brown and Atkinson, 2001, Berland et al., 2018). In two (5%) of the papers, funding of experimentation approach was adopted to enhance their ability to respond quickly and appropriately to changing market conditions (Zorzetti et al., 2022, Soe and Drechsler, 2018), which aligns with the agile principle of responding to changes over following a strict plan (Fowler and Highsmith, 2001). In Sirkiä and Laanti (2015), the Step-wise funding approach was adopted to minimize risk by postponing decisions, allowing the organization to make more informed strategic investment plans based on customer feedback. The motivation for adopting alternative and contracting approaches are further elaborated below.

### 4.2.2 Motivations for adopting alternative contracting approaches

The experiences of adopting Target Cost in this review were synthesized from the construction industry, manufacturing firms, housing sector, and software development. Organizations were found to adopt Target costing aiming for better cost management and higher profit margins (Afonso et al., 2008, Hattami et al., 2020, Baharudin and Jusoh, 2015, Cooper and Chew, 1996). Also, organizations appeared to favor Target Cost because of the need for more transparency and trust to foster stronger collaboration between clients and contractors (Badenfelt, 2008, Eckfeldt et al., 2005, Afonso et al., 2008, Cooper and Chew, 1996, Badenfelt, 2007).

The motivations for adopting Payment per sprint/milestone, Time and materials (T&M), and Two-phase contracts were primarily from software development projects. The findings showed that organizations adopted the Payment per Sprint/milestone approach aiming to minimize risk for both the client and contractor by promoting a more equitable sharing of risks (Zijdemans and Stettina, 2014). The Two-phase contract was adopted to balance the known structure of traditional approaches in one phase with the flexibility of Agile in another (Zijdemans and Stettina, 2014). T&M contracts were adopted to properly manage scope and value creation. For example, in Persson et al. (2022) where the client found that they were paying a lot of money for a product that was not satisfactory, so they adopted T&M. Table 4 summarizes the motivations for adopting the identified alternative funding and contracting approaches, along with the corresponding papers in which each motivation was identified.

*Table 4: Motivation for adopting alternative funding and contracting approaches*

| | Alternative approach | Motivation: Overall | Motivation: Specific | Paper |
|---|---|---|---|---|
| Alternative funding approaches | **Beyond budgeting** | Performance improvement | To enhance organizational agility, strategic alignment, and overall performance | (Ostergren and Stensaker, 2011) |
| | | | To be more flexible in operational planning and manage resources well (time and effort) | (Broccardo et al., 2016) |
| | | Effective response to changes | To enhance agility, innovation, and better alignment of management practices | (Bourmistrov and Kaarboe, 2013) |
| | | | To create a more flexible and responsive management accounting system that accommodates environmental uncertainties and reduces resource intensity | (Nikolaj Bukh and Sandalgaard, 2014) |
| | | | To offer a flexible, empowering, and effective approach to financial and organizational management | (Becker et al., 2020) |
| | | Cost management | To address the common issue of budget and schedule overruns in software projects | (Lohan et al., 2013) |
| | **Step-wise funding** | Effective response to changes | To make management and budgetary structures flexible and responsive | (Brander Brown and Atkinson, 2001, Ekholm and Wallin, 2011, Berland et al., 2018, Cao et al., 2013, Frow et al., 2010) |
| | | Performance improvement | To enable more focus on creativity, innovation, and strategic alignment | (Berland et al., 2018) |
| | | Risk minimization | To enable postponing decisions until a later time<br><br>To accommodate customer feedback and use it to facilitate strategic alignment | (Sirkiä and Laanti, 2015) |
| | **Funding of experimentation** | Effective response to changes | To enhance quick and appropriate responses to changing market conditions and user feedback | (Zorzetti et al., 2022, Soe and Drechsler, 2018) |
| Alternative contracting approaches | **Target cost** | Client-contractor collaboration | To foster stronger client-contractor partnerships | (Badenfelt, 2008) |
| | | | To enable the client to have full access to project costs and build trust between contractor and client | (Badenfelt, 2007) |
| | | | To establish contractual agreements that foster positive, productive client relationships | (Eckfeldt et al., 2005) |
| | | Performance improvement | It was seen as a competitive approach to managing strategies and operating at a profitable margin | (Baharudin and Jusoh, 2015) |
| | | | To remain competitive in global markets where consumers have more choices and demand better value. | (Cooper and Chew, 1996) |
| | | Cost management | To enable managing costs more effectively | (Jacomit and Granja, 2011, Badenfelt, 2008) |
| | | | To reduce cost and enhance efficiency in product development | (Dekker and Smidt, 2003) |
| | **Payment per sprint/milestone** | Risk minimization | To enable the sharing of risks fairly between client and contractor | (Thorup and Jensen, 2009, Zijdemans and Stettina, 2014) |
| | **Two-phase contract** | Effective response to changes | To leverage both the benefits of structure and flexibility from traditional and agile practices | (Zijdemans and Stettina, 2014) |
| | **Time and Materials contract** | Performance improvement | To enhance client satisfaction | (Persson et al., 2022) |
| | | Risk minimization | To ensure that the client does not pay more than required | (Persson et al., 2022) |

## 4.3 Benefits and challenges of adopting alternative funding and contracting approaches

### 4.3.1 Benefits of adopting alternative funding approaches

The specific benefits of each approach are presented below and summarized in Table 5, along with the corresponding papers in which the benefits were identified. An exception is the funding of experimentation, for which no reported benefits were reported in the reviewed papers.

*(a) Beyond budgeting*

As can be seen in Table 5, several findings indicate that Beyond Budgeting tends to increase customer satisfaction by empowering team members to be more innovative. Fixed-cost contracts often emphasize cost control, hindering innovation and strategic flexibility. In five (13%) of reviewed papers, the flexibility of Beyond budgeting was found to allow decisions to be postponed and made based on real-time feedback, thus creating excess time for teams to be innovative by focusing on other creative activities, which enhances customer satisfaction (Alsharari, 2020, Becker, 2014, Bourmistrov and Kaarboe, 2013, Ostergren and Stensaker, 2011, Lohan et al., 2013). Furthermore, flexibility facilitates collaboration as different departments work closely together continuously. It was found that agile teams worked closely with budgeting officers to constantly update the project budget based on agile sprint outcomes and feedback (Lohan et al., 2013).

The increased flexibility of Beyond Budgeting enables adjusting plans as new information is obtained. Findings from two (5%) of the papers showed that such flexibility aided organizations in managing risks by identifying and addressing issues early before they become large problems (Sirkiä and Laanti, 2015, Bourmistrov and Kaarboe, 2013). This supports existing studies, such as that of Nuottila et al. (2015) and Tian et al. (2015), highlighting that deferring decisions until more information is gathered helps avoid erroneous decisions, keeping options open longer, and allowing decisions to be made with fresher information. Additionally, four (11%) of the reviewed papers report that the flexibility provided by adopting Beyond Budgeting facilitates a more efficient and dynamic allocation by giving autonomy to managers to make decisions and reallocate resources based on changes that occur, which reduces the need for top management to micromanage every detail (Becker, 2014, Bourmistrov and Kaarboe, 2013, Ostergren and Stensaker, 2011, Lohan et al., 2013).

Existing studies, like that of de Waal (2005), show a positive correlation between adopting a Beyond Budgeting approach and better performance metrics. De Waal uses an example of Svenska Handelsbanken, showing that since it adopted the principles of Beyond Budgeting in 1970, it has consistently performed better than its European competitors on almost every performance indicator, such as return on equity, total shareholder return, earnings per share, cost-to-income ratio, and client satisfaction, suggesting that organizations that adopt Beyond Budgeting tend to outperform those using traditional approaches.

*(b) Step-wise funding*

Providing funds incrementally minimizes risk and ensures client satisfaction, allowing the client to change the project scope during development. A project lead in a software project discussed in one paper (3%) stated that incremental funding enabled the client only to fund the activities that deliver them value (Cao et al., 2013). Similarly, it was found in another paper (3%) that adopting a step-wise funding approach enabled the organization to focus on projects delivering the most value, leading to better alignment between investment and organizational goals (Thomas and Baker, 2008).

Two (5%) of reviewed papers showed that adopting Step-wise funding enabled frequent revisions and updates, making forecasting and resource allocation more efficient (Ekholm and Wallin, 2011, Frow et al., 2010). The flexibility for frequent budget revisions allowed funds to be reallocated from less to more critical areas without waiting for the lengthy approval processes typical of traditional budgeting. Moreover, the opportunity for frequent updates was found to promote teamwork and open communication, which ensured budget forecasts were regularly and accurately revised (Brander Brown and Atkinson, 2001). By doing so, cost was controlled without sacrificing agility (Ekholm and Wallin, 2011).

Findings from one paper (3%) suggested that Step-wise funding enhances flexibility by enabling managers to adapt strategies in response to changes (Berland et al., 2018). Existing studies like that of Jie Ng and Navaretnam (2019) suggest that the more adaptable an organization is to changes and feedback, the better it can meet user needs, leading to higher satisfaction and improved system usability. In another reviewed paper, Step-wise budgeting was found to expose organizational tensions, compelling discussions, and solutions (Berland et al., 2018). For example, it forced managers to engage with their teams to find a balance when there is tension between innovation and meeting organizational goals (Berland et al., 2018).

Findings from one paper (3%) further showed that adopting a Step-wise funding approach eliminated detailed cost-center planning, resulting in reduced bureaucracy and overhead (Sirkiä and Laanti, 2015). The organization simplified their cost center planning system, saving up to 50 man-days of planning work each month, thus freeing resources for other tasks. The review also found that the high flexibility achieved by Step-wise funding allowed organizations to respond effectively to market and customer feedback and allocate resources based on changing priorities without re-approvals or delays. This ensured that organizations focused on customer value and innovation rather than budget and fixed project scopes, leading to higher-quality outcomes (Sirkiä and Laanti, 2015). Similar to Bourne (2005), our findings highlight that Step-wise funding effectively addresses the limitations of traditional budgeting practices.

*(c) Pay As You Go (PAYG) based funding*

Findings from one paper (3%) showed that the PAYG approach was found to be appropriate for managing risk for both client and contractor (Cao et al., 2013). In Cao et al. (2013), a project manager who adopted the PAYG approach stated that this approach minimized the risk for the client to only one mini-project at a time instead of the entire project, and the contractors worked hard to ensure the client was satisfied so they could approve funding for the next mini-project.

### 4.3.2 Benefits of adopting alternative contracting approaches

*(a) Target cost*

Findings from two papers (5%) showed that Target Cost enhances creativity and innovation within design teams by fostering a culture of cost awareness and continuous improvement. The review showed that when a target cost is set, the teams tend to strive to meet the target cost and thus are motivated to explore ways to save costs and creative solutions that would enable them to achieve the desired cost objectives without compromising on product quality or customer satisfaction (Afonso et al., 2008,

Baharudin and Jusoh, 2015). By identifying cost targets and reduction strategies early, organizations can make informed design decisions aligned with financial goals, minimizing costly changes later in the process (Baharudin and Jusoh, 2015), .

Findings from three papers (8%) suggest that the ability to control costs through having a target cost was found to improve profit margins (Baharudin and Jusoh, 2015, Hattami et al., 2020, Cooper and Chew, 1996); however, the type of target cost contract matters. In a fixed-price type of contract, a common perception is that risk typically falls on the contractor. On the contrary, in a target cost contract, risk is perceived to be fairly shared between the contractor and client. Knowing that both would share the risks may incentivize both parties to work efficiently, ultimately improving project performance. Findings in Jørgensen (2017) show that when the contractor bears more risk (i.e., the contract has no upper limit on risk sharing), projects tend to perform worse than when there is an upper limit to risk sharing.

The flexibility of Target cost allows the contractor and client to adjust the scope during development. For example, in the Target cost contract adopted in Eckfeldt et al. (2005), changes in scope were allowed, and it was agreed beforehand that they could be categorized as either fixes (changes to meet functional requirements), clarifications (changes resulting from customer feedback), or enhancements (new stories that needed to be added but not originally agreed upon); and each category carried a different cost implication, giving the client ability to manage the budget and scope changes.

Additionally, findings from four (11%) of reviewed papers showed that sharing risks in Target costing promotes strong partnerships between clients and contractors, building trust, cooperation, and understanding (Ballard and Rybkowski, 2009, Baharudin and Jusoh, 2015, Eckfeldt et al., 2005, Cooper and Chew, 1996). When using target cost, the contractors were considered as business partners instead of contractors, which enabled better alignment with the client on various ways to reduce cost and complexity (Eckfeldt et al., 2005).

#### *(b) Payment per sprint/milestone*

Only one paper (3%) discusses benefits of Payment per sprint. Findings showed that Payment per sprint provides flexibility to adjust at the end of each sprint, and the client pays based on the actual value delivered at the end of the sprint. As discussed in Zijdemans and Stettina (2014), the contract begins with an agreement on a set number of sprints, and once the sprints are completed, the client can choose to purchase additional sprints. This approach is particularly beneficial for clients new to alternative contracting, as it allows them to start with a few sprints as a test. During this period, they can observe how the product backlog is created and managed, receive the first part of the system, and can decide whether to continue by purchasing more sprints based on the results. Payment per sprint was also found to minimize risk by offering fast payment to the contractor. In adopting this approach, the client only paid based on actual delivered work, and the supplier ensured that sprints were completed on time and met the client's needs (Zijdemans and Stettina, 2014).

Findings showed that adopting a contract that allows payments to be made in sprints or milestones enhances collaboration and trust between client and contractor. A collaborative environment also means risk is shared between both client and contractor as the client pays per sprint or increment delivered by the contractor, where the cost can be adjusted as required. In Zijdemans and Stettina (2014), a contractor shared their experience where payments were made after the completion of a sprint/milestone, and stated that it was the best collaboration climate they had encountered in 20 years (Zijdemans and Stettina, 2014). Such an environment allowed the client to receive timely assistance when encountering bugs, testing issues, or difficulties in setting up environments or clarifying requirements, and whenever the contractor faced technical challenges or struggled to understand requirements or business reasons, the client provided the required support (Zijdemans and Stettina, 2014).

#### *(c) Two-phase Contract*

Only one paper (3%) discusses benefits of Two-phase contract. Unlike a fixed-price contract, where one contract is used for both the design/planning phase and the development phase, the two-phase contract provides more opportunity to manage risks as the client has the opportunity to re-evaluate the project after Phase 1 before committing fully to the development costs in Phase 2. Findings further also showed that Two-phase contract enhanced collaboration as the client is involved throughout both phases, as clients can provide feedback and adjust priorities based on real-time progress (Zijdemans and Stettina, 2014).

#### *(d) Time & Materials (T&M)*

Findings from three (8%) of reviewed papers showed that projects that use T&M contracts have high flexibility to accommodate scope changes, which is important for agile projects as requirements are usually expected to change over time (Fink and Lichtenstein, 2014, Gopal and Koka, 2010, Jørgensen et al., 2017). T&M contracts were found to encourage frequent deliveries and feedback from the client, which promoted collaboration between the client and the contractor and resulted in having a high-quality product that met client satisfaction (Fink and Lichtenstein, 2014, Gopal and Koka, 2010, Jørgensen et al., 2017).

The T&M contract discussed in (Gopal and Koka, 2010), promoted continuous client-vendor collaboration, where clients provided feedback at regular intervals (e.g., during sprint reviews). Feedback at regular intervals ensured that the product met evolving expectations and aligned with business needs rather than adhering to a fixed initial set of requirements. In one reviewed paper by Thomas and Baker (2008), a client shared their experience of using T&M contracts by stating*, "I have experienced that budget management is much easier now. I sit among the developers and can see how they spend their time; every second week, I can prioritize what should be next… and I can decide to spend a specific amount, e.g., 10 million on maintenance".* Moreover, three (8%) of reviewed papers highlight the risk sharing in T&M contracts where the client pays for the actual work done, and the contractor has no penalties. Thus, the risk is shared, which would otherwise have all fallen to the contractor in a fixed contract setting (Fink and Lichtenstein, 2014, Gopal and Koka, 2010, Jørgensen et al., 2017).

*Table 5: Benefits of adopting the alternative approach*

| | Alternative approach | Benefits | | Papers |
|---|---|---|---|---|
| | | Overall | Specific | |
| Alternative Funding approaches | Beyond budgeting | Higher client satisfaction | Supports and emphasizes focus on high-quality performance, continuous improvement and stakeholder values | (Alsharari, 2020) |
| | | | Higher client satisfaction by empowering and engaging the workforce to be more innovative | (Becker, 2014, Bourmistrov and Kaarboe, 2013, Ostergren and Stensaker, 2011, Lohan et al., 2013) |
| | | Saves time and cost | Leads to a more timely and effective decision-making | (Bourmistrov and Kaarboe, 2013) |
| | | Efficient and dynamic use of resources | Facilitates efficient and dynamic use of resources | (Becker, 2014, Bourmistrov and Kaarboe, 2013, Ostergren and Stensaker, 2011, Lohan et al., 2013) |
| | | Promotes trust and collaboration | Promotes collaboration among teams | (Becker, 2014, Lohan et al., 2013) |
| | Step-wise funding | Higher client satisfaction | The client can fund only activities that deliver value | (Cao et al., 2013, Sirkiä and Laanti, 2015) |
| | | | Can lead to better alignment between investment and organizational goals. | (Thomas and Baker, 2008) |
| | | Efficient and dynamic use of resources | Organizations can flexibly allocate resources based on changing priorities and needs | (Ekholm and Wallin, 2011, Frow et al., 2010), (Berland et al., 2018, Sirkiä and Laanti, 2015) |
| | | Promotes trust and collaboration | Encourages teamwork and open communication | (Brander Brown and Atkinson, 2001) |
| | | Saves time and cost | Enables cost control without sacrificing agility | (Ekholm and Wallin, 2011) |
| | | | Ensures that financial resources are always supporting current strategic priorities | (Frow et al., 2010) |
| | | | Exposes organizational tensions, forcing them to be addressed on a timely manner | (Berland et al., 2018) |
| | | | Organizations can reduce bureaucracy and overhead | (Sirkiä and Laanti, 2015) |
| | Pay -As-You-Go (PAYG) based funding | Higher client satisfaction | Minimizes risk by focusing on only one phase at a time instead of the entire project and ensures the client is satisfied before moving to next phase | (Cao et al., 2013) |
| Alternative contracting approaches | Target cost | Saves time and cost | By identifying cost targets and reduction strategies upfront, the company could make informed design decisions that aligned with financial goals, reducing the need for costly changes later in the development process | (Baharudin and Jusoh, 2015) |
| | | | Improve profit margins through proper cost control and product design optimization | (Hattami et al., 2020, Baharudin and Jusoh, 2015, Cooper and Chew, 1996) |
| | | Promotes trust and collaboration | Fosters a partnership mentality between contractors and clients, leading to a high level of trust, cooperation and understanding | (Ballard and Rybkowski, 2009, Eckfeldt et al., 2005, Cooper and Chew, 1996, Baharudin and Jusoh, 2015) |
| | | Higher client satisfaction | Ensures that product quality and features closely match market expectations, thus leading to higher customer satisfaction. | (Afonso et al., 2008, Baharudin and Jusoh, 2015) |
| | | Efficient and dynamic use of resources | Allows for significant adaptability in project scope and requirements, accommodating the iterative and dynamic nature of Agile development | (Eckfeldt et al., 2005) |
| | Payment per sprint /milestone | Promotes trust and collaboration | Increases transparency between the client and contractor and minimizes risk for the contractor | (Zijdemans and Stettina, 2014) |
| | | Enhances cost management | Provides higher flexibility to the client as the customer can buy more sprints | (Zijdemans and Stettina, 2014) |
| | Two-phase contract | Promotes trust and collaboration | Risk is minimized as the contract is distributed in phases and adopts the project needs and uncertainty. | (Zijdemans and Stettina, 2014) |
| | **Times & Materials (T&M)** | Promotes trust and collaboration | Encourage closer collaboration between clients and contractors<br><br>Minimizes risk as the client only pays for actual time and materials used | (Fink and Lichtenstein, 2014, Gopal and Koka, 2010, Jørgensen et al., 2017, Persson et al., 2022) |
| | | Higher client satisfaction | Allows for changes in requirements and adaptability during project execution, which can lead to better alignment with client needs | (Jørgensen et al., 2017, Persson et al., 2022) |

### 4.3.3 Challenges of adopting alternative funding approaches

The specific challenges of each approach are presented below and summarized in Table 6, along with the corresponding papers in which the benefits were identified. An exception is the funding of experimentation and Pay-As-You-Go-based funding, for which no challenges were reported in the reviewed papers.

*(a) Beyond budgeting*

The review found that adopting Beyond Budgeting may be accompanied by several challenges. In four (8%) papers, organizations reported the need for significant cultural, organizational, and mindset shifts (Alsharari, 2020, Bourmistrov and Kaarboe, 2013, Ostergren and Stensaker, 2011, Lohan et al., 2013, Aksom, 2019). For example, an organization that adopted beyond budgeting had to also change its organizational structure In Aksom (2019), it was found that viewing Beyond Budgeting purely as a financial issue rather than an organizational change and people not knowing the core principles of Beyond budgeting, hindered its successful adoption. Existing research also shows that successful adoption of Beyond budgeting requires both managers and employees to actively move out of their comfort zones (traditional approaches) and into stretch zones (alternative approaches) (Heupel and Schmitz, 2015). Consistent with de Waal (2005), the review found in two papers (5%) that strong leadership with a good understanding of Beyond Budgeting principles is essential to guide and support its successful implementation (Becker, 2014, Bourmistrov and Kaarboe, 2013).

Due to the need for a change in control and management structure when adopting Beyond Budgeting, it was reported in two papers (5%) that organizations may encounter resistance from individuals accustomed to traditional structures (Becker, 2014, Bourmistrov and Kaarboe, 2013). While Beyond Budgeting increases flexibility, it was reported in three papers (5%) to add complexity to the budgeting process (Becker, 2014, Ostergren and Stensaker, 2011, Lohan et al., 2013). In Lohan et al. (2013), it was reported that the increased complexity led to financial mismanagement as the project team excessively adjusted their budget allocations beyond what was justified, which then required additional corrective actions to realign the budget.

Another challenge reported in six (16%) of reviewed papers is the fear among management of losing control over organizational processes (Alsharari, 2020, Lohan et al., 2013, Ostergren and Stensaker, 2011, Aksom, 2019), (Becker, 2014, Bourmistrov and Kaarboe, 2013). The review also found that organizations may easily revert to old approaches if circumstances change, such as when leaders who championed Beyond Budgeting become replaced by new leaders (Becker, 2014). According to Becker (2014), new leaders may feel the need to reintroduce traditional approaches to regain control. Additionally, two (5%) of the papers reported that the lack of structure in Beyond Budgeting poses difficulty for management to evaluate and manage performance (Becker, 2014, Ostergren and Stensaker, 2011). Additionally, the review showed that Beyond Budgeting relies heavily on continuous collaboration across all project levels, which can lead to miscommunication and delays in budget adjustments (Lohan et al., 2013). Despite these challenges, the effort to adopt Beyond Budgeting is said to be worthwhile, as its benefits significantly outweigh the difficulties (Heupel and Schmitz, 2015).

The review found that without proper preparation, organizations may approach Beyond Budgeting too cautiously, which may jeopardize its successful adoption (Aksom, 2019), therefore to facilitate the successful adoption of Beyond budgeting, substantial learning about its principles and significant adjustments to organizational processes is required (Becker, 2014, Bourmistrov and Kaarboe, 2013). These findings also align with Tian et al. (2015), who highlight that adopting Beyond Budgeting demands additional time for individuals to become familiar with and adjust to the new system.

### *(b) Step-wise funding*

Experiences from two papers (5%) reported that that step-wise funding system requires significantly more administrative resources. For example, in Ekholm and Wallin (2011), it was reported that a dedicated accounting team working closely with management was required in adopting this approach. Since Step-wise funding requires a high level of flexibility, Frow et al. (2010) reported that it necessitated ongoing data collection and analysis, which placed a heavy burden on teams and diverted resources from other critical tasks

Adopting Step-wise funding in an organization requires a significant cultural shift. The organization studied in one paper had relied for a long time on traditional approaches, and they reported experiencing resistance from individuals accustomed to strict control who felt uneasy with the transparency and flexibility of Step-wise funding approach. And after adopting Step-wise funding, the organization struggled to provide accurate long-term cost estimates and had had to accept rough long-term projections and short-term estimates for more precise planning (Sirkiä and Laanti, 2015).

Two papers (5%) reported the potential for "analysis paralysis" as a challenge that accompanies the adoption of Step-wise funding (Ekholm and Wallin, 2011, Frow et al., 2010). Analysis paralysis occurs when the presence of numerous alternatives leads to overthinking and an inability to make decisions. For example, in Ekholm and Wallin (2011), it was reported that the introduction of flexible budgeting led to the finance team spending more time revising forecasts and budgets rather than focusing on strategic initiatives. Thus, suggesting that a high level of flexibility can result in delays in budgeting decisions.

Three papers (8%) reported internal conflicts as a challenge in adoption of Step-wise funding (Berland et al., 2018, Ekholm and Wallin, 2011, Frow et al., 2010). Such an approach allows for frequent budget adjustments, which may lead to potential internal conflicts. It was reported in Ekholm and Wallin (2011) that frequent refining and adjusting of action plans can lead to tensions between sticking to existing strategies or adjusting to new developments. In Berland et al. (2018), managers were restricted to only five strategic actions, which forced them to prioritize between which tasks were feasible and which ones they could realistically manage. Thus, although a Step-wise funding approach may encourage creativity, it may lead to confusion with regard to the prioritization of ideas to pursue and which to leave as they are.

### 4.3.4 Challenges of adopting alternative contracting approaches

The specific challenges of each approach are presented below and summarized in Table 6, along with the corresponding papers in which the challenges were reported. An exception is the Funding of experimentation and Two-Phase contract, which were not explicitly described in the reviewed papers.

*(a) Target cost*

Three (8%) of papers report on the challenges of adopting Target cost. Findings showed that adopting Target cost involved a shift from a traditional to a more collaborative and value-driven approach (Ballard and Rybkowski, 2009, Eckfeldt et al., 2005). However, establishing a collaborative environment between client and contractor posed challenges, particularly in finding the right balance of risk-sharing. Trust is an important element in building collaboration in target costing. One paper reported that the client may increase control if they sense unreliability from the contractor, which can harm collaboration and elevate the risk (Badenfelt, 2007). The reviewed papers reported that the successful adoption of target cost required continuous communication and effort (Ballard and Rybkowski, 2009, Eckfeldt et al., 2005). In Eckfeldt et al. (2005), a contractor who wanted to adopt Target cost for agile projects had to put extra effort into convincing a client unfamiliar with the approach by explaining its benefits and alignment with agile principles. The contractor promised continuous communication, negotiation, and clear guidelines to align changes with the project's strategic goals. In addition to constant communication and effort between client and contractor, adopting Target costing required continuous effort and communication between all team members throughout the project's design phase Eckfeldt et al. (2005).

*b) Payment per sprint/milestone*

Two papers (5%) report challenges of adopting payment per sprint/milestone (Thorup and Jensen, 2009, Zijdemans and Stettina, 2014). The adoption of a collaborative agile contract was found to rely heavily on a strong foundation of trust and cooperation between contractor and client (Thorup and Jensen, 2009). The review also found that more effort is required from the client to ensure success when using target costing. Initially, the client invested effort in clearly defining expectations and responsibilities in the contract. Throughout development, the client monitored progress closely and made constant adjustments and negotiations necessary to align with the evolving project scope and goals, which increased the risk of scope creep (Thorup and Jensen, 2009). The review also found that paying per sprint may lead to a short-term focus as contractors may focus on delivering as much functionality as possible within the agreed sprints, which may compromise the long-term quality of the software and potentially result in technical debt or poor software architecture (Zijdemans and Stettina, 2014).

*c) Time & Materials (T&M)*

Since the contractor is paid hourly in T&M contracts, there may not be the incentive to finish tasks fast, which then requires the client to be actively involved in monitoring to avoid significant budget increases. Three papers (8%) report challenges for adopting T&M contracts (Jørgensen et al., 2017, Gopal and Koka, 2010, Jørgensen and Grov, 2021). Findings from Jørgensen and Grov (2021) showed that although the cost of T&M contracts may be lower than fixed price contracts, a T&M contract requires additional effort from the client to monitor the progress of the project. Findings from Gopal and Koka (2010), where the use of fixed-price and T&M contracts was compared, suggested that because T&M contracts are paid based on effort (work hours) and the risk falls on the client, the contractor may lack the motivation to achieve high quality. The review further found that contractors had difficulty convincing clients to adopt T&M contracts because they feared such contracts might be too open-ended and have the potential for unchecked costs (Eckfeldt et al., 2005).

*Table 6: Challenges of adopting the alternative funding and contracting approaches*

| | Alternative approach | Challenges | | Papers |
|---|---|---|---|---|
| | | General | Specific | |
| Alternative Funding approaches | Beyond Budgeting | Requires a shift in mindset | Initial resistance from people due to lack of structure as that provided by fixed budgets | (Becker, 2014, Bourmistrov and Kaarboe, 2013) |
| | | | Requires substantial learning and adjusting by the organization, hence needs strong leadership to provide guidance | (Alsharari, 2020, Lohan et al., 2013, Ostergren and Stensaker, 2011, Aksom, 2019), (Becker, 2014, Bourmistrov and Kaarboe, 2013) |
| | | Internal conflicts | Potential for financial mismanagement due to the increased complexity and flexibility of the budgeting process | (Lohan et al., 2013) |
| | | Requires additional resources (effort and time) | With no fixed budgets, evaluating and managing performance becomes more difficult | (Becker, 2014, Ostergren and Stensaker, 2011) |
| | Step-wise funding | Analysis paralysis | Requires significantly more administrative resources and advanced IT support than a fixed-budget model | (Ekholm and Wallin, 2011) |
| | | | Over-flexibility may result in delays in decision-making | (Ekholm and Wallin, 2011, Frow et al., 2010) |
| | | Internal conflicts | It is complex to manage a flexible budgeting system effectively and may lead to internal conflicts due to frequent budget adjustments | (Ekholm and Wallin, 2011, Frow et al., 2010) |
| | | | Can create tensions in prioritizing, innovation, and strategy | (Berland et al., 2018) |
| | | Requires additional effort | The need for ongoing data collection and analysis places a heavy burden on teams, sometimes diverting resources from other critical tasks | (Frow et al., 2010) |
| | | | It makes it difficult to track costs while maintaining flexibility | (Sirkiä and Laanti, 2015) |
| | | Requires a shift in mindset | Requires a significant shift in culture | (Sirkiä and Laanti, 2015) |
| Alternative Contracting approaches | Target cost | Requires a shift in mindset | Requires a shift in mindset from traditional cost management practices to a more collaborative and value-driven approach | (Ballard and Rybkowski, 2009) |
| | | | Necessitates continuous communication and negotiation with clear guidelines and mutual understanding to ensure that changes remain within projects' strategic goals | (Eckfeldt et al., 2005) |
| | | Requires additional resources (effort) | Determining the appropriate balance of risk sharing between the contractor and the client can be challenging | (Eckfeldt et al., 2005, Ballard and Rybkowski, 2009) |
| | | | Target costing is meeting-intensive and requires continuous effort from all team members throughout the project's design phase | (Ballard and Rybkowski, 2009) |
| | Payment per sprint/ milestone | Requires additional resources (effort) | Requires continuous adjustments and negotiations to align with the evolving project scope and goals. | (Thorup and Jensen, 2009) |
| | | Risk of scope creep | Given the iterative and flexible approach to defining project scope, there is a significant risk of scope creep | (Thorup and Jensen, 2009) |
| | | Requires a shift in mindset | Heavily relies on a strong foundation of trust and cooperative behavior between the customer and the supplier | (Thorup and Jensen, 2009) |
| | | Short term vision | May lead to short-term focus | (Zijdemans and Stettina, 2014) |
| | Times & Materials (T&M) | Requires additional effort | More effort is required from the client to monitor progress and ensure proper use of resources to avoid cost overruns and minimize risk | (Jørgensen et al., 2017, Gopal and Koka, 2010, Jørgensen and Grov, 2021) |

# 5. DISCUSSION AND LIMITATIONS

## 5.1 Experiences of adopting alternative funding and contracting approaches

The overall experiences with adopting alternative funding and contracting approaches are positive. Across the reported contexts, alternative funding and contracting approaches resulted in similar benefits related to time and cost savings, higher user satisfaction, better resource utilization, and improved trust and collaboration between client and contractor.

Although it has been evident that the adoption of alternative funding and contracting approaches leads to significant benefits, it is challenging to adopt such approaches in organizations that are used to doing things the traditional way (Boehm and Turner, 2005). According to our findings, these challenges are related to the need for change in the mindset of the people involved, additional resources in terms of effort, time, and finances, internal conflicts due to increased complexity, scope creep, and analysis paralysis. Existing studies support our finding that the change of mindset is one of the key factors if an organization is to adopt alternative funding and contracting approaches successfully (Chen et al., 2016, Cooper and Sommer, 2016).

These challenges are not surprising, as introducing new ways naturally comes with difficulties and some resistance (Obina and Adenike, 2022). Similarly, replacing traditional with alternative funding and contracting approaches will require significant effort from the organization. Heupel and Schmitz (2015) emphasize that the effort is worth it as the benefits of adopting alternative approaches far outweigh the challenges. To encourage organizations to adopt alternative approaches, Heupel and Schmitz (2015) further suggest that regulators and leaders could attempt to enforce policies, introduce industry standards, or promote frameworks that favor their adoption. Without such encouragement, reluctance may continue, and organizations may miss out on the benefits.

## 5.2 Adoption of alternative approaches across domains

Alternative funding and contracting approaches have been adopted across several domains. The extent of their adoption and the diversity of domains in which they have been reported differ significantly. However, each of the funding and contracting approaches identified by this review have at least one reported experience of being used within the software domain.

Among the identified contracting approaches, Target Cost appears to be the most widely adopted, with the highest number of reported experiences spanning across multiple domains. Such domains include software, manufacturing, construction, oil and gas, housing, agriculture, and hospitality. The diverse widespread across domains aligns with existing studies that have found target costing to be adopted in high-tech consumer goods and project-based sectors like manufacturing and construction (Zimina et al., 2012). This finding is not surprising as Target costing approach has a long history of almost a century (Feil et al., 2004). Organizations are therefore more confident in adopting this approach due to a continual experience of positive experiences spanning many years. This review suggests that Target cost contracting is the most adopted alternative approach with the potential to benefit various domains beyond software.

Beyond Budgeting also demonstrates significant representation across multiple domains, including software. Studies that specify domains include software, banking, retail, manufacturing, oil and gas, and agriculture. However, it is worth noting that most of the studies in this review presenting experiences of using Beyond Budgeting did not specify a particular industry but rather focused on overarching business, managerial, or accounting perspectives. Some papers mention that the experience is from the public sector or family and non-family-owned firms without being specific on the industrial domains. This may suggest that Beyond Budgeting is perceived more as a general financial management strategy irrespective of a specific domain.

In contrast, Step-wise funding was found to be predominantly adopted within software development, with some adoption in manufacturing specifically new product development. However, there is a notable gap in cross-domain experiences. In our review, we found that in the step-wise funding approach, the release of the next round of funding depends on the performance of the previous stage. Our finding may suggest that this approach is best suited for contexts where a project's continuation depends on feedback from its progress. This characteristic aligns well with agile software development and new product development in manufacturing, as both involve innovation, novelty, and uncertainty, which require frequent revisions. Thus, this funding model supports iterative development and continuous improvement which is particularly effective in agile environments. However, its adoption in industries with more structured, linear and predictable processes appears to be less common, highlighting the need for further exploration across different domains.

Several approaches such as Time & Materials (T&M), Pay-As-You-Go (PAYG), Funding of experimentation, payment per sprint and Two-phase contracts, were found to be exclusively reported in the software domain. Our review did not find evidence of their application in other industries, suggesting a research gap that needs further exploration or limited reported experiences adoption outside software development.

The diversity of experiences reported across these funding and contracting approaches highlights their potential adoption beyond the software domain. Even though each industry operates in a distinct context, cross-domain insights are valuable for understanding the effectiveness of these approaches (Nuottila et al., 2015). The more positive experiences that are documented across domains, the better we can understand their usefulness, prerequisites, and best practices for successful adoption.

## 5.3 Which is the "best" alternative funding or contracting approach for agile projects?

In selecting which alternative approach to adopt, organizations should be aware that no one approach is inherently better than another, but rather, it is the context that plays a major role. Approaches like Beyond Budgeting, Time and Materials (T&M), and Target Cost have more research-based evidence available across various sectors and a longer history. However, even lesser-known approaches can yield positive outcomes if applied in the right context. The key is selecting an approach that fits the specific needs of the organization. For example, the Step-wise approach termed "Adaptive finance and control" was developed by integrating Lean, Agile, and Beyond Budgeting principles. The organization that adopted this approach considered its own specific context and needs and developed this tailored approach allowing it to reap the benefits of flexibility from the integration of the three principles (Sirkiä and Laanti, 2015). The most important task when selecting an alternative approach is to ensure issues such as (i) the level of understanding of those involved, (ii) the readiness of the leaders in the organization to make the change, and (iii) the ability of the organizational systems to integrate new approaches, are considered throughout the adoption process.

Based on the above results, we believe it may be useful for an organization to assess its context and maturity and choose a funding and/or contracting approach that provides a good fit and does not involve very big process changes per step. For example, it may not be necessary for first-time adopters to abandon traditional approaches completely. Instead, they may start with minor modifications and gradually incorporate flexible practices to complement traditional ones. This will allow them to develop customized approaches that work best for their specific context. This is supported by Cardoş (2014) who, in his review of new budgeting trends, highlights that alternative approaches are not standardized solutions and, instead, organizations should customize an approach to fit their internal systems, taking into account their unique culture, structure, and infrastructure.

## 5.4 The implementation of alternative funding or contracting approaches

Our findings suggest that adopting alternative funding and contracting approaches can have characteristics ranging from evolutionary to revolutionary approaches. An example of adopting an evolutionary approach can be replacing a Fixed cost contract with a Fixed price and negotiable scope contract, while an example of adopting a revolutionary approach can be transitioning directly from traditional budgeting to Beyond Budgeting or Step-wise funding. Although the revolutionary adoption appears more appealing for organizations, the experiences from the review show that it may not always be possible or necessary to make a total revolution and instead, opting for a hybrid approach may be the optimal way (Ekholm and Wallin, 2011, Cao et al., 2013, Aksom, 2019). Nikolaj Bukh and Sandalgaard (2014) reported that organizations with specific structural characteristics struggled with the complexities of Beyond Budgeting, and instead of adopting it right away, they adopted a hybrid approach by combining both traditional and Beyond Budgeting principles.

A strong positive correlation was found between the perceived usefulness of a mix of both traditional and alternative approaches, suggesting they should be seen as complements, not rivals (Ekholm and Wallin, 2011). According to Cao et al. (2013), traditional funding approaches can be adapted to support the unique characteristics of agile software development projects rather than introducing completely new approaches. Aksom (2019) reports a preference for organizations adapting aspects of Beyond Budgeting to fit within existing organizational frameworks rather than fully adopting the Beyond Budgeting approach. These findings support existing studies suggesting there may be a preference for organizations to integrate some principles of alternative approaches without entirely abandoning traditional approaches as they allow the integration of both elements of agile and traditional, enabling organizations to capitalize on the strengths of both (Cooper and Sommer, 2018, Nguyen et al., 2018).

## 5.5 Public sector lagging behind

The public sector remains especially reluctant to adopt alternative funding and contracting approaches, but it may benefit from learning from private sector experiences (Ocampo et al., 2021). Our review reports only a few examples of public sector adoption, which may be caused by the public sector's strong emphasis on cost control, bureaucracy, and rigid systems (Nuottila et al., 2016, Baxter et al., 2023, Bogdanova et al., 2020). Additionally, the experiences from the public sector in this review have been limited to Target Cost, Beyond Budgeting, and Funding of experimentation. Target Cost and Beyond Budgeting are older, well-known approaches that have consistently shown positive outcomes across sectors, which may have encouraged the public sector to be more open to adopting them.

For other approaches that have not existed long and have primarily been created for agile projects, such as Step-wise funding and Pay-As-You-Go based funding, their experiences in this review were found only in the private sector. Despite the emphasized need for flexibility in agile projects and the benefits that come with them, we observe the reluctance of the public sector to adopt alternative funding and contracting approaches due to the desire not to lose control. As reported in the introduction, in 2019, the government of New South Wales in Australia introduced an alternative funding approach for digitalization projects called the Digital Restart Fund. Despite being in its early years, it has reported success stories by making funds available quicker and in a step-wise manner. The Digital Restart Fund shows that alternative approaches can be implemented successfully at the government level.

This review suggests that many organizations may already be using alternative approaches under different names that are not widely recognized. For example, Dekker and Smidt (2003) report that more than half of the organizations surveyed in their study used Target Cost, but they referred to it by different terminologies. Similarly, organizations have also developed approaches tailored to the specific needs of the contractors and clients during negotiations. Thorup and Jensen (2009), for example, describe Payment per sprint contract, which they termed a Collaborative agile contract. Given these observations, it is likely that more organizations in the private sector, or even the public sector, are adopting alternative funding and contracting approaches with positive outcomes. However, due to the infancy of this topic, there remains a lack of standardized terminology across sectors, industries, and contexts, which leaves such experiences fragmented and difficult to synthesize. To address this challenge and encourage the public sector to explore and adopt alternative funding and contracting approaches, more research is required to continue synthesizing existing experiences.

## 5.6 Limitations

There are several limitations of the systematic literature review presented in this paper. Before starting the review, we conducted a preliminary study to identify alternative funding and contracting approaches in existing literature. The findings informed on the keywords used in our search string. However, due to the lack of standardized terminology in alternative funding and contracting, the review cannot claim to have exhausted all alternative contracting and funding approaches for agile projects. This poses a threat to construct validity as variations in terminology may have excluded some relevant studies. To mitigate this threat, we searched across multiple databases (Scopus, Web of Science and Google scholar), applied forward and backward snowballing and sought expert recommendation to expand our coverage.

Most likely, there is a risk of publication bias in the reviewed papers, as organizations with negative experiences may be less likely to report on their experience with alternative funding or contracting approaches. This presents potential threat to external validity. Therefore, our results should not be interpreted as representative but mainly as documenting the existence of successful uses of alternative funding and contracting approaches.

We only included peer-reviewed and published publications in English language, excluding all unpublished and grey literature. This means we may have overlooked relevant experiences documented in non-peer-reviewed organizational reports. This may, in particular, be the case for governmental reports claiming the use of agile funding approaches[6]. This presents a threat to external validity as it may have limited the generalizability of our findings.

The results on benefits and challenges of the use of the alternative approaches and our interpretation of them require identification of the outcome effect of one particular approach (the funding or contracting approach) in a very complex context with many other components, such as the type of problem solved and the competence of the involved resources. This presents a threat to internal validity, as it is difficult to determine whether the benefits were caused solely by the alternative approach. Therefore, in many cases, the results should be considered mainly correlational, not necessarily causal, and based on perceptions rather than actual causal analyses. Fortunately, in several instances, there are good reasons to believe there was at least a positive contribution from adopting the funding or contracting approach. As an illustration, several papers (see sections 1 and 2) find a positive impact of more flexible specifications on agile software development. This is consistent with our finding that funding and contracting approaches allowing for more flexibility in scope and cost have a positive effect on the outcomes.

To strengthen conclusion validity, we ensured that our review process was well-documented and replicable. We followed established research guidelines, i.e., (Kitchenham, 2004) and took measures to reduce the risk of selection bias. To address the risk of bias in data extraction, two researchers independently reviewed the selected papers, and we formulated a set of questions that we followed to guide the reviewing process. Finally, the synthesis of findings was conducted collaboratively to ensure consistency in interpretation.

# 6. CONCLUSION

This study aims to synthesize practical experiences of adopting alternative funding and contracting approaches that are potentially relevant for use in the context of agile software development. Through a systematic literature review process, we identified four alternative funding approaches: Beyond budgeting, Step-wise funding, Pay-As-You-Go based funding, and funding of experimentation. Additionally, we identified four alternative contracting approaches: Target costing, Payment per sprint, Two-phase contract, and Time and Materials (T&M). The study highlights both the positive and negative experiences organizations encounter when transitioning from traditional to alternative approaches.

The identified alternative funding and contracting approaches have shown significant potential to improve efficiency, client satisfaction, resource utilization, and risk management by enabling organizations to adapt readily to evolving project demands and market conditions. The positive experiences presented in this study can encourage organizations, specifically the public sector, that have been slower to adopt these approaches, to explore and adopt them. However, adoption also introduces challenges, such as increased resource requirements, potential decision paralysis from scope creep, and cultural shifts necessitating strong change management strategies. Awareness of these challenges can help organizations prepare to adopt these alternative approaches successfully.

[6] Such reports exist, for example, for Australia and their “Digital Restart Fund (www.digital.nsw.gov.au/funding).

While alternative approaches offer substantial advantages, each organization should select an approach that meets its unique needs. The study emphasizes the importance of a tailored approach, i.e., considering each organization's specific context, structure, and readiness, as no single approach universally fits all contexts. This review confirms that, with proper implementation, alternative funding and contracting approaches can enhance greater agility and alignment with agile principles. Thus, to ensure that more organizations benefit from these approaches, they are encouraged to adopt alternative approaches progressively, starting with hybrid ones that combine traditional and agile approaches and then moving toward fully agile approaches as they build familiarity with using them.

### 6.1 Future research

This study may be the first to synthesize real-life experiences of adopting alternative funding and contracting approaches relevant for agile software development projects. As part of the review, we identified several areas for further research. Our review indicates that the private sector is more open to adopting alternative funding and contracting approaches than the public sector. Future research could explore the policies, standards, or frameworks that might encourage public organizations to adopt such approaches.

The experiences in this review are drawn from completed projects retrospectively from organizations that adopted alternative approaches, with the exception of one paper that performed a longitudinal study following a project for two years from initiation to completion. Future studies could conduct more longitudinal studies tracking the projects from initiation to completion could provide deeper insights into the experiences of adopting alternative funding and contracting approaches over time. This review did not consider contextual characteristics, such as project size and complexity when analyzing the benefits and challenges connected with the use of alternative funding and contracting approaches. Future studies could explore which approaches are most effective for small versus large agile projects, or low- versus high-complexity agile projects, and potentially other contextual factors. Such knowledge may help to increase understanding for organizations seeking to adopt alternative approaches and guide them to select a suitable approach for their projects.

### Declaration of interest

The authors declare they have no known competing financial interests or personal relationships that could have influenced the work.

# 8. Appendices

*Appendix i: Journal and Conference Ranking Assessment*

| Reference | Journal/ Conference | Impact factor (2023) | JOURQUAL (2024) | ABDC (2022) | SJR score 2023 | CORE conference ranking |
|---|---|---|---|---|---|---|
| (Dekker and Smidt, 2003) | International Journal of production economics | 2.86 | 3 | A | Q1 | N/A |
| (Jørgensen and Grov, 2021) | Information and Software Technology | 3.8 | - | A | Q1 | N/A |
| (Lohan et al., 2013) | 2012 International Conference on Information Systems Development | - | N/A | N/A | N/A | A |
| (Jacomit and Granja, 2011) | Architectural Engineering and Design Management | 1.3 | - | - | Q1 | N/A |
| (Alsharari, 2020) | International journal of public sector management | 2.5 | 1 | - | Q2 | N/A |
| (Cao et al., 2013) | European Journal of Information Systems | 7.3 | 4 | A* | Q1 | N/A |
| (Sirkiä and Laanti, 2015) | 48th International Conference on System Sciences | N/A | N/A | N/A | N/A | A |
| (Soe and Drechsler, 2018) | Government Information Quarterly | 7.8 | - | A | Q1 | N/A |
| (Nikolaj Bukh and Sandalgaard, 2014) | Journal of Accounting & Organizational Change | 2.4 | - | B | Q1 | N/A |
| (Brander Brown and Atkinson, 2001) | International Journal of Contemporary Hospitality Management | 9.1 | 3 | A | Q1 | N/A |
| (Broccardo et al., 2016) | World Review of Entrepreneurship, Management and Sustainable Development | 0.98 | - | C | Q3 | N/A |
| (Zijdemans and Stettina, 2014) | 15th International Conference on Agile Processes in Software Engineering and Extreme Programming | N/A | N/A | N/A | N/A | B |
| (Thorup and Jensen, 2009) | 2009 Agile Conference | N/A | N/A | N/A | N/A | * |
| (Frow et al., 2010) | Accounting, Organizations and Society | 3.6 | 4* | A* | Q1 | N/A |
| (Jørgensen et al., 2017) | International Journal of Project Management | 7.4 | 2 | A | Q1 | N/A |
| (Thomas and Baker, 2008) | 2008 Agile Conference | N/A | N/A | N/A | N/A | * |
| (Berland et al., 2018) | Journal of Applied Accounting Research | 3.9 | 2 | - | Q1 | N/A |
| (Bourmistrov and Kaarboe, 2013) | Management Accounting Research | 4.2 | 2 | - | Q1 | N/A |
| (Zorzetti et al., 2022) | Information and Software Technology | 3.8 | - | A | Q1 | N/A |
| (Ramakrishnan, 2009) | Issues in Informing Science & Information Technology[7] | - | - | - | - | N/A |
| (Ostergren and Stensaker, 2011) | European Accounting Review | 2.5 | 3 | A* | Q1 | N/A |
| (Aksom, 2019) | International Journal of Management Practice | 0.78 | - | C | Q3 | N/A |
| (Persson et al., 2022) | International Journal of Business Information Systems | 1.03 | - | C | Q3 | N/A |
| (Ballard and Rybkowski, 2009) | Construction Research Congress | N/A | N/A | N/A | N/A | * |
| (Al-Hattami et al., 2020) | International Journal of Business Excellence | 0.97 | - | - | Q3 | N/A |
| (Eckfeldt et al., 2005) | Agile Development Conference | N/A | N/A | N/A | N/A | * |
| (Jørgensen, 2017) | The 10th International Workshop on Cooperative and Human Aspects of Software Engineering (CHASE) | N/A | N/A | N/A | N/A | * |
| (Cooper and Chew, 1996) | Harvard Business Review | 9.1 | 3 | A | Q3 | N/A |
| (Baharudin and Jusoh, 2015) | Procedia - Social and Behavioral Sciences[8] | N/A | N/A | N/A | N/A | * |
| (Ax et al., 2008) | International Journal of Production Economics | 9.8 | 3 | A | Q1 | N/A |
| (Ekholm and Wallin, 2011) | Journal of Business Finance & Accounting | 2.2 | 3 | A* | Q1 | N/A |
| (Afonso et al., 2008) | International Journal of Production Economics | 9.8 | 3 | A | Q1 | N/A |
| (Becker et al., 2020) | Contemporary Accounting Research | 3.2 | 4 | A* | Q1 | N/A |
| (Gopal and Koka, 2010) | Decision sciences | 2.8 | 3 | A* | Q1 | N/A |
| (Badenfelt, 2008) | Engineering, Construction and Architectural Management | 3.6 | - | A | Q1 | N/A |
| (Badenfelt, 2007) | The 23rd ARCOM Annual Conference | N/A | N/A | N/A | N/A | * |
| (Becker, 2014) | European Accounting Review | 2.5 | 3 | - | Q1 | N/A |
| (Fink and Lichtenstein, 2014) | ACM SIGMIS Database: the DATABASE for Advances in Information Systems | - | - | A | Q2 | N/A |

[7] While this journal is not indexed in Scopus or Web of Science, nor ranked in the ABDC or Harzing lists, it maintains a strong h-index of 36 and g-index of 57, indicating a consistent research impact, though not at the highest level. These metrics are comparable to those of Q2/Q3 journals, reflecting its steady academic influence.

[8] Procedia - Social and Behavioral Sciences was discontinued in 2019, but Scopus during its publication years it was indexed in Scopus.

* Conferences that do not have a CORE ranking but whose proceedings are indexed in Scopus, indicating that they meet certain quality standards. Being indexed in Scopus signifies that the conference is reputable, as its papers have undergone review and are recognized as meaningful contributions to the field of research.

## *Appendix ii: Data extraction form*

| | **Data extracted from each paper** | **Response options expected, among others** |
|---|---|---|
| 1 | Paper Title | |
| 2 | Authors | |
| 3 | Source of Publication (Tick appropriate) | ☐ Journal ☐ Conference |
| 4 | Journal/Conference Name | |
| 5 | Year of Publication | |
| 6 | Location of Study | |
| 7 | Unit Studied (Tick appropriate) | ☐ Project ☐ Organization ☐ Other: _____ |
| 8 | Sector (Tick appropriate) | ☐ Public ☐ Private ☐ Other: _____ |
| 9 | Research Method (Tick appropriate) | ☐ Qualitative ☐ Quantitative ☐ Mixed |
| 10 | Data Collection Method (Tick appropriate) | ☐ Interviews ☐ Survey/Questionnaire ☐ Workshop ☐ Other: _____ |
| 11 | Sample Size | |
| 12 | Context of the study | |
| 13 | Comment (if any) on Traditional approaches | |
| 13 | Alternative Funding or Contracting approach Discussed | ☐ Incremental Funding ☐ Beyond Budgeting ☐ Target Cost ☐ Pay-as-you-go (PAYG) ☐ Funding of experimentation ☐ Other: _____ |
| 14 | Motivation for adopting the alternative Funding/ Contracting approach | |
| 15 | Reported Benefits of adopting the alternative Funding/ Contracting approach) | |
| 16 | Reported Challenges/Shortcomings of adopting alternative Funding/ Contracting approach) | |
| 17 | Additional Comments (if any) | |

## *Appendix iii: Papers included in the review*

| **Title of the paper** | **Authors** | **Year** | **Published** | **Location** | **Unit studied** | **Sector** | **Data collection method** | **Sample size** | **Alternative approach** |
|---|---|---|---|---|---|---|---|---|---|
| A survey of the adoption and use of target costing in Dutch firms | Henri Dekker and Peter Smidt | 2003 | Journal | Netherlands | Organization | Private | Survey | 43 responses | Target cost |
| A field experiment on trial sourcing and the effect of contract types on outsourced software development | Magne Jørgensen, and Jon Grov | 2021 | Journal | Not specified | Project | Private | Controlled field experiment | 16 software development projects (36 developers) | Time and Materials |
| A performance management model for agile information systems development teams | Garry Lohan, Michael Lang, and Kieran Conboy | 2012 | Conference | Not specified | Project | Private | Interviews | 19 interviews from two multinational companies | Beyond budgeting |
| An Investigation into the Adoption of Target Costing on Brazilian Public Social Housing Projects | Ana Mitsuko Jacomit and Ariovaldo Denis Granja | 2011 | Journal | Brazil | Organization | Private | Interviews, documents and archival records | 1 case organization (interview number unspecified) | Target cost |
| Accounting changes and beyond budgeting principles (BBP) in the public sector: Institutional isomorphism | Nizar Mohammad Alsharari | 2020 | Journal | Jordan, UAE | Organization | Public | Interviews, documents and observations | 80 interviews | Beyond budgeting |
| Adapting funding processes for agile IT projects: an empirical investigation | Lan Cao, Kannan Mohan, Balasubramaniam Ramesh, and Sumantra Sarkar | 2013 | Journal | Not specified | Project | Private | Interviews, observations, and documents | Respondents from 12 different projects | Step-wise funding and Pay-As-You-Go |
| Adaptive Finance and Control: Combining Lean Agile and Beyond Budgeting for Financial and Organizational Flexibility | Rami Sirkiä and Maarit Laanti | 2015 | Conference | Finland | Organization | Private | Authors' direct experiences with the transformation | A case study approach - 1 organization | Step-wise funding |
| Agile local governments: Experimentation before implementation | Ralf-Martin Soea and Wolfgang Drechsler | 2017 | Journal | Finland and Estonia | Project | Public | Interviews, documents, observation | 5 organizations | Funding of experimentation |
| Beyond Budgeting and Change: A Case Study | Niels Sandalgaard and Per Nikolaj Bukh | 2014 | Journal | Denmark | Organization | Private | Interviews, reports, website information | 15 interviews with managers and top management | Beyond budgeting |
| Budgeting in the information age: a fresh approach | Jackie Brander Brown Helen Atkinson | 2001 | Journal | US | Organization | Private | Case study - observations, documents | NA | Step-wise funding |
| Budgeting process: an Italian survey in family and non-family firms | Laura Broccardo, Elisa Giacosa, Francesca Culasso and Alberto Ferraris | 2016 | Journal | Italy | Organization | Mix | Survey | 48 medium-large sized companies | Beyond budgeting |
| Contracting in Agile Software Projects: State of Art and How to Understand It | Shi Hao Zijdemans and Christoph Johann Stettina | 2014 | Conference | Netherlands | Organization | Private | Interviews and Workshop | A 45 min workshop with 8 participants and 5 interviews | Two-phase contract and Payment per sprint |
| Collaborative Agile Contracts | Lars Thorup and Bent Jensen | 2009 | Conference | Denmark | Project | Private | Author's experiences from 2 projects | 2 projects | Payment per sprint |
| Continuous Budgeting: Reconciling Budget Flexibility with Budgetary Control" by | Natalie Frow, David Marginson, and Stuart Ogden | 2009 | Journal | UK | Organization | Private | Interviews, documents, and observations | 31 interviews | Step-wise funding |
| Direct and indirect connections between type of contract and software project outcome | Magne Jørgensen, Parastoo Mohagheghi, and Stein Grimstad | 2017 | Journal | Norway | Project | Public | Survey, interviews, documents and reports | 3467 T&M projects and 107 Interviews | Time and Materials |
| Establishing an Agile Portfolio to Align IT Investments with Business Needs | Joseph C. Thomas and Steven W. Baker | 2008 | Conference | USA | Organization | Private | Case study | Not specified | Step-wise funding |
| Exposing organizational tensions with a non-traditional budgeting system | Nicolas Berland, Emer Curtis and Samuel Sponem | 2018 | Journal | France | Organization | Private | Interviews | 33 interviews, documents and audit reports from 15 Business Units | Step-wise funding |
| From comfort to stretch zones: A field study of two multinational companies applying "beyond budgeting" ideas | Anatoli Bourmistrov and Katarina Kaarbøe | 2013 | Journal | Sweden | Organization | Private | Interviews, observations, presentations, meetings, and documents | Multiple case study – 2 companies (oil and Gas and Telecom) | Beyond budgeting |
| Improving Agile Software Development using User-Centered Design and Lean Startup | Maximilian Zorzetti, Ingrid Signoretti, Larissa Salerno, Sabrina Marczak and Ricardo Bastos | 2022 | Journal | Not specified | Project | Private | Interviews, focus group discussions and workshops | 15 interviews, 8 FGDs, 1 workshop | Funding of experimentation |
| Innovation and Scaling up Agile Software Engineering Projects | Sita Ramakrishnan | 2009 | Journal | Australia | Project | Public | Interviews | 15 individual evaluations, Observations | Funding of experimentation |
| Management Control without Budgets: A Field Study of 'Beyond Budgeting' in Practice | Katarina Østergren and Inger Stensaker | 2010 | Journal | Norway | Organization | Private | Interviews, and documents | 20 interviews | Beyond budgeting |
| Managerial understanding and attitudes towards beyond budgeting in Ukraine | Herman Aksom | 2019 | Journal | Ukraine | Organization | Private | Interviews | 11 interviews | Beyond budgeting |
| Outsourcing Agile Projects with Time-and-Material Contracts: A Trust-Control Duality Perspective | John Stouby Persson, Lene Pries-Heje and Lars Mathiassen | 2022 | Journal | Finland | Project | Public | Interviews | 1 organization followed from contracting to completion (2 years) | Time and Materials |
| Overcoming the Hurdle of First Cost: Action Research in Target Costing | Glenn Ballard and Zofia K. Rybkowski | 2009 | Conference | US | Project | Private | Action research | Action research – a single case project | Target cost |
| Reducing costs in manufacturing firms by using target costing technique | Hamood Mohd. Al-Hattami, Jawahar D. Kabra and Murlidhar A. Lokhande | 2020 | Journal | Yemen | Organization | Private | Interviews, and observation | 3 people from management (exact number of interviews unspecified) | Target cost |
| Selling Agile: Target-Cost Contracts | Bruce Eckfeldt, Rex Madden and John Horowitz | 2005 | Conference | US | Project | Private | Authors' direct experiences | 1 project | Target cost |
| Software development contracts: The impact of the provider's risk of financial loss on project success | Magne Jørgensen | 2017 | Conference | Norway | Project | Mix | Survey | 24 responses | Target cost |
| Target Costing: Let Customers Drive Pricing | Robin Cooper and W. Bruce Chew | 1996 | Journal | Japan | Organization | Private | Case study - 2 organizations | 2 organizations | Target cost |
| Target Cost Management (TCM): a case study of an automotive company | Norhafiza Baharudina and Ruzita Jusoh | 2015 | Journal | Malaysia | Organization | Private | Interviews | 24 personnel from several managerial levels | Target cost |
| The Impact of Competition and Uncertainty on the Adoption of Target Costing | Christian Ax, Jan Greve, and Ulf Nilsson | 2008 | Journal | Sweden | Organization | Private | Survey | 57 responses | Target cost |
| The Impact of Uncertainty and Strategy on the Perceived Usefulness of Fixed and Flexible Budgets | Bo-Goran Ekholm and Jan Wallin | 2011 | Journal | Scandinavian countries | Organization | Private | Survey | 342 responses from top CEOs and CFOs of large manufacturing companies | Step-wise funding |
| The influence of time-to-market and target costing in the new product development success | Paulo Afonso, Manuel Nunes, Antonio Paisana, and Ana Braga | 2008 | Journal | Portugal | Organization | Private | Survey | 82 responses | Target cost |
| The Interplay of Core and Peripheral Actors in the Trajectory of an Accounting Innovation: Insights from Beyond Budgeting | Sebastian D. Becker, Martin Messner, and Utz Schäffer | 2020 | Journal | Europe | Organization | Private | Interviews, 23 presentations, workshops, conferences, and meetings | 25 interviews | Beyond budgeting |
| The role of contracts on quality and returns to quality in offshore software development outsourcing | Anandasivam Gopal and Balaji R. Koka | 2010 | Journal | India | Project | Private | Interviews, and Survey | 100 software projects (40 projects used T&M contract) | Time and Materials |
| The selection of sharing ratios in target cost contracts | Ulrika Badenfelt | 2007 | Journal | Sweden | Project | Private | Interviews | 16 interviews (8 clients and 8 contractors) | Target cost |
| Trust and control in the early phases of Target Cost Contracts | Ulrika Badenfelt | 2007 | Conference | Sweden | Project | Private | Interviews | 4 interviews (2 contractors and 2 clients) | Target cost |
| When Organisations Deinstitutionalise Control Practices: A Multiple-Case Study of Budget Abandonment | Sebastian D. Becker | 2014 | Journal | Europe | Organization | Private | Interviews | 22 interviews from 4 organizations | Beyond budgeting |
| Why project size matters for contract choice in software development outsourcing | Lior Fink and Yossi Lichtenstein | 2014 | Journal | Europe | Project | Private | Survey | 267 software development projects | Time and Materials |